\documentstyle[11pt,amssymb,epsf,cite]{article}

\makeatletter
\@addtoreset{equation}{section}
\makeatother

\def\ben{\begin{equation}}
\def\een{\end{equation}}

\let\a=\alpha    
    
  \let\n=\nu

\let\C=\Chi

\def\nn{\nonumber} \def\bd{\begin{document}} \def\ed{\end{document}}
\def\ds{\documentstyle} \let\fr=\frac \let\bl=\bigl \let\br=\bigr
\let\Br=\Bigr \let\Bl=\Bigl
\let\bm=\bibitem
\let\na=\nabla
\let\pa=\partial \let\ov=\overline
\newcommand{\be}{\begin{equation}}
\newcommand{\ee}{\end{equation}}
\def\ba{\begin{array}}
\def\ea{\end{array}}
\def\ft#1#2{{\textstyle{{\scriptstyle #1}\over {\scriptstyle #2}}}}
\def\fft#1#2{{#1 \over #2}}
\def\del{\partial}
\def\vp{\varphi}
\def\sst#1{{\scriptscriptstyle #1}}
\def\oneone{\rlap 1\mkern4mu{\rm l}}
\def\td{\tilde}
\def\wtd{\widetilde}
\def\ie{\rm i.e.\ }
\def\dalemb#1#2{{\vbox{\hrule height .#2pt
        \hbox{\vrule width.#2pt height#1pt \kern#1pt
                \vrule width.#2pt}
        \hrule height.#2pt}}}
\def\square{\mathord{\dalemb{6.8}{7}\hbox{\hskip1pt}}}
\newcommand{\ho}[1]{$\, ^{#1}$}
\newcommand{\hoch}[1]{$\, ^{#1}$}
\newcommand{\bea}{\begin{eqnarray}}
\newcommand{\eea}{\end{eqnarray}}
\newcommand{\ra}{\rightarrow}
\newcommand{\lra}{\longrightarrow}
\newcommand{\Lra}{\Leftrightarrow}
\newcommand{\ap}{\alpha^\prime}
\newcommand{\bp}{\tilde \beta^\prime}
\newcommand{\tr}{{\rm tr} }
\newcommand{\Tr}{{\rm Tr} }
\def\0{{\sst{(0)}}}
\def\1{{\sst{(1)}}}
\def\2{{\sst{(2)}}}
\def\3{{\sst{(3)}}}
\def\4{{\sst{(4)}}}
\def\5{{\sst{(5)}}}
\def\6{{\sst{(6)}}}
\def\7{{\sst{(7)}}}
\def\8{{\sst{(8)}}}
\def\n{{\sst{(n)}}}
\def\cA{{{\cal A}}}
\def\cF{{{\cal F}}}
\def\tV{\widetilde V}
\def\tW{\widetilde W}
\def\tH{\widetilde H}
\def\tE{\widetilde E}
\def\tF{\widetilde F}
\def\tA{\widetilde A}
\def\im{{{\rm i}}}
\def\tY{{{\wtd Y}}}
\def\ep{{\epsilon}}
\def\vep{{\varepsilon}}
\def\R{\rlap{\rm I}\mkern3mu{\rm R}}
\def\bD{{{\bar D}}}
\def\cD{{\cal D}}
\def\R{\rlap{\rm I}\mkern3mu{\rm R}}
\def\bD{{{\bar D}}}
\def\R{{{\Bbb R}}}
\def\C{{{\Bbb C}}}
\def\H{{{\Bbb H}}}
\def\CP{{{\Bbb C}{\Bbb P}}}
\def\RP{{{\Bbb R}{\Bbb P}}}
\def\Z{{{\Bbb Z}}}
\def\bA{{{\Bbb A}}}
\def\bB{{{\Bbb B}}}
\def\bC{{{\Bbb C}}}
\def\bD{{{\Bbb D}}}
\def\bZ{{{\Bbb Z}}}
\def\Re{{{\frak{Re}}}}
\def\Im{{{\frak{Im}}}}
\def\cosec{{\,\hbox{cosec}\,}}
\def\tX{{{\wtd X}}}

\newcommand{\tamphys}{\it Center for Theoretical Physics,
Texas A\&M University, College Station, TX 77843, USA}
\newcommand{\umich}{\it Michigan Center for Theoretical Physics,
University of Michigan\\ Ann Arbor, MI 48109, USA}
\newcommand{\upenn}{\it Department of Physics and Astronomy,
University of Pennsylvania\\ Philadelphia,  PA 19104, USA}
\newcommand{\SISSA}{\it  SISSA-ISAS and INFN, Sezione di Trieste\\
Via Beirut 2-4, I-34013, Trieste, Italy}

\newcommand{\ihp}{\it Institut Henri Poincar\'e\\
  11 rue Pierre et Marie Curie, F 75231 Paris Cedex 05}

\newcommand{\damtp}{\it DAMTP, Centre for Mathematical Sciences,
 Cambridge University\\ Wilberforce Road, Cambridge CB3 OWA, UK}
\newcommand{\itp}{\it Institute for Theoretical Physics, University of
California\\ Santa Barbara, CA 93106, USA}

\newcommand{\auth}{M. Cveti\v{c}\hoch{\dagger}, G.W. Gibbons\hoch{\sharp}
and C.N. Pope\hoch{\ddagger}}

\begin{document}
\begin{flushright}
\hfill{DAMTP-2003-75\ \ \ MIFP-03-23\ \ \ UPR-1064-T}\\
\hfill{hep-th/0401151}
\end{flushright}


\begin{center}
{ \large {\Large\bf Ghost-Free de Sitter Supergravities as Consistent
Reductions of String and M-theory}}

\vspace{20pt}
\auth

\vspace{3pt}
{\hoch{\dagger}\upenn}

\vspace{3pt}


\vspace{3pt}
{\hoch{\sharp}\damtp}

\vspace{3pt}
{\hoch{\ddagger}\tamphys}

\vspace{30pt}

\underline{ABSTRACT}
\end{center}

     We study properties of supergravity theories with non-compact
gaugings, and their higher-dimensional interpretations via consistent
reductions on the inhomogeneous non-compact hyperboloidal spaces
${\cal H}^{p,q}$. The gauged supergravities are free of ghosts,
despite the non-compactness of the gauge groups.  We give a general
discussion of the existence of stationary points in the scalar
potentials  of such supergravities.  These are of
interest since they can be associated with de Sitter vacuum
configurations.  We give explicit results for consistent reductions
on ${\cal H}^{p,q}$ in various examples, derived from analytic
continuation of previously-known consistent sphere reductions.
In addition  we also consider black hole and cosmological
solutions,  for specific examples of non-compact gaugings in $D=4,5,7$.

{\vfill\leftline{}\vfill \vskip 5pt \footnoterule {\footnotesize
\hoch{\dagger} Research supported in part by DOE grant
DE-FG02-95ER40893, NATO grant 97061, NSF grant $\phantom{xxxxx}$
INTO-03585 and the Fay R. and Eugene L.  Langberg Chair. \vskip
-12pt} \vskip 14pt {\footnotesize \hoch{\star} Research supported
in full by DOE grant DE-FG02-95ER40899 \vskip -12pt} \vskip 14pt
{\footnotesize  \hoch{\ddagger} Research supported in part by DOE
grant DE-FG03-95ER40917.\vskip  -12pt}}

\pagebreak
\setcounter{page}{1}

\tableofcontents
\addtocontents{toc}{\protect\setcounter{tocdepth}{3}}
\vfill\eject

\section{Introduction}

    Over the past few years important progress
\cite{spheresa,domainw,vann,consistent,d4gauge,spheresc} has been made
in understanding the full non-linear structure of certain Kaluza-Klein
sphere reductions of string and M-theory, leading to gauged
supergravities with maximal supersymmetry in lower dimensions. These
reductions, also referred to as Pauli reductions \cite{Pauli}, are
consistent only for specific supergravity theories compactified on
spheres of specific dimension \cite{consistent,Pauli}.  In particular,
they lead to gauged supergravity theories with anti-de Sitter
vacua having a {\it negative
cosmological constant}, or to dilatonic vacua corresponding to
domain wall solutions with a potential of the type $-|\lambda|
\, e^{\phi}$ \cite{domainw}.

    On the other hand the origin of de-Sitter vacua arising from
consistent reductions of string and M-theory is less well studied.  It
is known that there exist gauged supergravities with non-compact gauge
groups, which can be obtained from the usual compact gaugings by means
of appropriate analytic continuations.  These were extensively studied
in \cite{Hull,hullwarn,GibbonsHull}.  With the more recent advances in
understanding the higher-dimensional origins of the compact gauged
supergravities via consistent sphere reductions, it is therefore
worthwhile re-examining the non-compact gaugings, with a view to
studying their higher-dimensional origins from string or M-theory.

   The essential features of the geometrical structures involved in
the reductions to non-compact gauged supergravities can be summarised
as follows.  For the compact gauged theories, notably in $D=7$, 5 and
4 dimensions, one makes reductions of $D=11$ supergravity on $S^4$,
type IIB supergravity on $S^5$ or $D=11$ supergravity on $S^7$
respectively.  In many cases, the general structure of the internal
$S^n$ metric is
\be
ds^2 = T^{-1}_{AB}\, d\mu^A\, d\mu^B\,,\label{rn1met}
\ee
where $T^{-1}_{AB}$ is a matrix of scalar fields, and
$\mu^A$ are coordinates on $\R^{n+1}$,  subject to the constraint
\be
\delta_{AB}\, \mu^A\, \mu^B = 1\label{snconstr}
\ee
which restricts the $\mu^A$ to lie on the sphere.
In each case there exists a ground state where the
lower-dimensional scalar fields vanish, corresponding to $T_{AB}=\delta_{AB}$,
and the internal metric becomes that of the
round sphere $S^n$,
as
\be
ds^2 =\delta_{AB}\,  d\mu^A\, d\mu^B\,.\label{roundmet}
\ee
Here, of course, both (\ref{snconstr}) and (\ref{roundmet}) are invariant
under $SO(n+1)$, and thus the internal manifold in the vacuum state has
the $SO(n+1)$ isometry of the round sphere $S^n$.

   In the reduction to the non-compact gauging, the constraint
(\ref{snconstr}) is replaced by
\be
\eta^{\phantom{\Sigma_\Sigma}}_{AB}\, \mu^A\, \mu^B = 1\,, \label{hpqconstr}
\ee
where $\eta_{AB}=\hbox{diag}(1,1,\ldots, 1, -1, -1,\ldots -1)$, with
$p$ eigenvalues $+1$, and $q=n+1-p$ eigenvalues $-1$.  However, the
``trivial'' scalar configuration (which may, or may not, be a stationary
point of the scalar potential) is still given by $T_{AB}=\delta_{AB}$.
Thus the internal metric in the trivial-scalar configuration is
given by
\be
ds^2 = \delta_{AB}\, d\mu^A\, d\mu^B\,,\qquad
\eta^{\phantom{\Sigma_\Sigma}}_{AB}\, \mu^A\, \mu^B=1\,.\label{hpqmet}
\ee
The first equation is invariant under $SO(p+q)$ while the second is
invariant under $SO(p,q)$, and so the metric is invariant under the
common subgroup, $SO(p)\times SO(q)$.  The spaces described by these
metrics are hyperboloidal, and are designated by ${\cal H}^{p,q}$
\cite{Hull,GibbonsHull}.  Note that the metrics are always positive
definite.  When $p$ and $q$ are both non-zero, the space ${\cal
H}^{p,q}$ is {\it non-compact}, and furthermore it is {\it
inhomogeneous} (\ie it is not a coset space).  Particular reductions
of supergravity theories on the non-compact hyperboloidal spaces
${\cal H}^{p,q}$ yield theories in lower dimensions that may have
vacuum solutions with positive cosmological constant.   The scalar
fields $T_{AB}$ play an essential role in these theories, in
ensuring that all the lower-dimensional gauge fields (as well as the
scalars themselves) have standard positive-energy kinetic terms,
despite the occurrence of non-compact gauge
groups.\footnote{There are also reductions
of the so-called *-theories \cite{*Theory} on the hyperboloidal spaces
with positive definite metric (see e.g., \cite{LPP} and references
therein) may provide examples of stable de-Sitter vacua of
anti-de Sitter vacua (see, e.g., \cite{BehrndtCvetic} and references
therein).  Note however that in these cases the *-theory already
suffers from ghost-fields and the non-linear Kaluza-Klein ansatz for
the $p$-form field strengths involves complex values.}

    It is worth emphasising the distinction between the nature
of the reductions we are considering here, which involve the inhomogeneous
non-compact spaces ${\cal H}^{p,q}$, and reductions involving 
non-compact spaces that have non-compact isometry groups.
In our reductions, the fiducial metric defined by (\ref{hpqmet})
necessarily has a {\it compact} isometry group, and this lies at the
heart of why one obtains lower-dimensional theories with no ghost-like
gauge fields associated with ``wrong-sign'' kinetic terms. The full
non-compact $SO(p,q)$ gauge group is always spontaneously broken in any
solution, and in fact the residual unbroken gauge group is always
compact.  By contrast, if one performs a reduction on a space with
a {\it non-compact} isometry group, such as the homogeneous hyperbolic
plane ${\bf H}^2 = SL(2,\R)/O(2)$, there will always be ghost-like
gauge fields associated with the non-compact generators of the 
isometry group.  This is because, in a linearised analysis of small
fluctuations around the vacuum solution, the Yang-Mills field strengths
in the kinetic terms will be contracted with the indefinite-signature
Cartan-Killing metric of the non-compact isometry 
group.\footnote{It should be stressed, therefore, that it is the
signature of the Cartan-Killing metric of the isometry group, and {\it not}
the signature of the metric on the internal space itself, that governs
the signs of the gauge-field kinetic terms.}  

    If we consider the example
of 2-dimensional non-compact spaces, the two contrasting situations 
can be illustrated by considering ${\cal H}^{2,1}$ and ${\bf H}^2$, 
defined by
\bea
{\cal H}^{2,1}:\qquad&& 
ds^2 = d\mu_1^2 + d\mu_2^2 + d\mu_3^2\,,\qquad \mu_1^2 + \mu_2^2
-\mu_3^2 =1\,,\nn\\
{\bf H}^2:\qquad &&
d\td s^2 = d\mu_1^2 + d\mu_2^2 - d\mu_3^2\,,\qquad \mu_1^2 + \mu_2^2
-\mu_3^2 =1\,.
\eea
Both of these metrics have positive-definite signature, but ${\cal
H}^{2,1}$ is inhomogeneous, with the isometry group $O(2)$, while
${\bf H}^2$ is homogeneous, with isometry group $O(2,1)$.  The ${\cal
H}^{2,1}$ metric provides a basis for reducing to give an $O(2,1)$
gauged ghost-free supergravity with (at most) a surviving $O(2)$ gauge
group in the vacuum.  By contrast, the ${\bf H}^2$ metric could yield
an $O(2,1)$ gauged theory which would have indefinite-signature
kinetic terms for the gauge fields, and a vacuum with a surviving
$O(2,1)$ gauge group. (The question of whether one can find {\it
consistent} reductions, for which the massive Kaluza-Klein towers can
be set to zero, is a more subtle one. However, this is quite distinct
from the present question of whether or not the kinetic terms for the
gauge fields have the correct sign for ghost-freedom.)
 
   The purpose of this paper is to analyse possible consistent
reductions of string and M-theory on hyperboloidal spaces ${\cal
H}^{p,q}$, and the properties of the vacuum solutions for the
resulting theories in lower dimensions. The analysis is facilitated by
the fact that we can obtain these reductions as analytic continuation
of sphere reductions whose consistent non-linear Kaluza-Klein
Ans\"atze have been extensively studied
\cite{spheresa,vann,d4gauge,spheresc}.  

     Our discussions will focus on the bosonic sectors of the
supergravity theories, and so we will not generally be explicitly
addressing the important question of whether the reductions and
truncations we study are also compatible with supersymmetry.  However,
the general results on non-compact gaugings in works such as
\cite{Hull,hullwarn} show that the necessary supersymmetric
completions of the bosonic sectors do indeed exist, and this provides 
compelling evidence to support the idea that our consistent reductions 
can be extended to include the fermionic sectors.

    The paper is organised as follows. In section \ref{extremasec} we
analyse the extrema of the scalar field potentials that generically
arise in theories obtained from a reduction of string and M-theory on
hyperboloidal spaces ${\cal H}^{p,q}$, and for completeness we analyse
the extrema of the potentials arising from reductions on spheres
$S^{p-1}$.  At such extrema one can truncate the theory to fixed
values of the scalar fields. In the case of more that one extremum (as
is the case for compact reductions) one can also address the
properties of renormalisation group flows (in the dual field
theories) interpolating between such (non-positive) stationary points
of the scalar potential.  Note that potentials arising from
hyperboloidal reductions always turn out to have positive definite
extrema, while in the sphere reductions the potentials are negative
at the extrema. In the case of $S^{p-1}$ with $p\le 3$ there is only one
extremum, while for $p\ge 4$ there is a second extremum with a larger
negative value of the potential.

   In section 3 we discuss the non-linear Kaluza-Klein Ans\"atze for
Pauli reductions on both compact (spheres $S^{p+q-1}$) and
non-compact (hyperboloidal ${\cal H}^{p,q}$) spaces, by employing a
description of scalar fields in terms of the vielbeine on the scalar
coset manifold $SL(n,\R )/SO(n)$ ($n=p+q$).  It should be emphasised that
although the non-compact reductions are derived from the compact ones
by analytic continuation, the internal manifolds ${\cal H}^{p,q}$ are
inhomogeneous, even though the original spheres $S^{p+q+1}$ are
homogeneous spaces. The explicit example that
is described in detail is that of $(p,q)=(4,4)$, a reduction of
11-dimensional supergravity on $S^7$ and ${\cal H}^{4,4}$.  Other
consistent examples of sphere reductions can analogously be discussed
in the context of hyperboloidal reductions as well.

   In section 4 we focus on the study of de Sitter supergravity in
four-dimensions, beginning with the $N=4$ theory, obtained as a Pauli
reduction of 11-dimensional supergravity on the hyperboloidal space
${\cal H}^{4,4}$ space.  In particular, we employ an analytic
continuation to derive this reduction from the corresponding
consistent reduction on $S^{7}$. In addition, we discuss a truncation
of this theory to $N=2$ de Sitter supergravity, which makes contact
with a recent result in the literature \cite{lupo}.  We also consider
more general $N=2$ theories in four dimensions, obtaining via analytic
continuations examples corresponding to reductions on ${\cal H}^{4,4}$
and ${\cal H}^{6,2}$.  In section 5, we consider examples of de Sitter
type gauged supergravities in five and seven dimensions.

   In section 6, we present examples of charged black hole and cosmological solutions
   for specific examples of non-compact gaugings in $D=4,5,7$.  These  solutions
   are related to the  corresponding multi-charged black holes of AdS gauged
   supergravities.  Concluding  remarks are given in section 7.

\section{Extrema of Scalar Potentials}\label{extremasec}

    In this section, we shall focus our attention on scalar potentials
of the form
\be
V = 2 T_{ij}\, T_{ij} - (T_{ii})^2\,,
\ee
where $T_{ij}$ is a symmetric matrix of scalar fields.  There are
several examples in gauged supergravities where potentials of this type
arise from dimensional reductions on spheres.  Commonly, but not
always, the scalar matrix is unimodular.  To encompass also those
cases where it is not restricted to be unimodular, it is useful
to extract the determinant by defining
\be
   T_{ij} = \Phi\, \wtd T_{ij}\,,\qquad \hbox{where}\quad
  \det(\wtd T_{ij})=1\,,
\ee
and thus we have
\be
V = \Phi^2\, \wtd V\,,\qquad \wtd V \equiv
2 \wtd T_{ij}\, \wtd T_{ij} - (\wtd T_{ii})^2\,,  \label{poten}
\ee

     Our study will begin by considering the stationary points of
$\wtd V$.  One application of these results will be for discussing the
circumstances under which one can perform a consistent truncation of
the scalar fields in the associated gauged supergravity.  In order to
be able to truncate the theory to fixed values of the scalars $\wtd
T_{ij}$, the necessary condition,
dictated  by  the
equations of motion, is that these fixed values
correspond to a
stationary point of their potential.

   The scalar matrix $\wtd T_{ij}$ is conveniently parameterised in terms of
a scalar vielbein $\wtd\Pi_A{}^i$, in terms of which one has
\be
\wtd T_{ij} = \wtd\Pi_i^{-1\, A}\, \wtd\Pi_j^{-1\, B}\, \eta_{AB}\,.
\ee
We shall consider the situation of scalars associated with a gauging
of $SO(p,q)$, for which we shall have the $SO(p,q)$ invariant metric
\be
\eta_{AB}= {\rm diag}\, (+1,+1,\ldots,+1,-1,-1,\ldots,-1)\,,
\ee
with $p$ plus signs and $q$ minus signs.  The vielbein $\wtd\Pi_A{}^i$
parameterises the scalar coset manifold $SL(n,\R)/SO(n)$, where
$n=p+q$.  Note that the denominator group $SO(p+q)$ is always compact
regardless of whether the gauge group $SO(p,q)$ is compact or
non-compact.  Thus $i,j$ are $SO(p+q)$ indices, raised and lowered
with $\delta_{ij}$, while $A,B$ are $SO(p,q)$ indices, raised and
lowered with $\eta_{AB}$.

   In order to study the extrema of the potential $\wtd V$, it is
convenient to perform local transformations to diagonalise the
scalar vielbein, implying that we can write
\be
{\tilde T}_{ij}={\rm diag}\, (X_1,X_2,\dots,X_p,-X_{\bar 1},-X_{\bar
2},\cdots,-X_{\bar q})\ ,
\ee
where $X_a$ and $X_{\bar a}$ are all positive, subject to the
unimodularity constraint:
\be
\prod_{a=1}^p X_a \prod_{{\bar a}=1}^q X_{\bar a}=1\,.\label{cons}
\ee
Note that we must have $q=2r$, where $r$ is an integer, in order to have
$\det(\wtd T_{ij})=+1$ with positive $X_a$ and $X_{\bar a}$.

    The potential $\wtd V$ now takes the form:
\be
\wtd V= 2\sum_{a=1}^p X_a^2 +2 \sum_{{\bar a}=1}^q X_{\bar a}^2
  - \Big(\sum_{a=1}^p X_a -\sum_{\bar a=1}^q X_{\bar a}\Big)^2\,.\label{pot}
\ee

   Without loss of generality we may assume that $p\ge q$, since
if $q$ exceeded $p$ we could simply redefine our notion of what is
a time-like and what a space-like direction.  (The overall sign of
$\wtd T_{ij}$ plays no role in the analysis.)

   The discussion at this stage divides into two cases, depending on
whether $q=2r=0$ (the compact case), or $q=2r\ge 2$ (the non-compact
case).  We shall begin by considering the non-compact case.

\subsection{Non-compact Case}\label{noncompsec}

   In this subsection we shall consider non-compact cases, where
$q=2r\ge 2$.

   The extrema of $\wtd V$ can be determined  by introducing a Lagrange
multiplier to enforce the constraint (\ref{cons}), and defining
\be
S = \wtd V + \lambda\, \Big( \prod_{a=1}^p X_a\, \prod_{\bar a=1}^q X_{\bar a}
-1\Big)\,.
\ee
The equations following from requiring $S$ to be stationary under
the variations of the $X_a$ and $X_{\bar a}$ imply
\bea
X_a^2-2\, {\sigma}_- \, X_a +\ft14 \lambda&=&0\, ,  \ \  a=1,\cdots , p \,,
\label{exta}\\
X_{\bar a}^2 + 2\, {\sigma}_-\,  X_{\bar a} +\ft14 \lambda&=&0\, ,\ \
b=1,\cdots , q \,,\label{ext}
\eea
where $\sigma_-\equiv {\textstyle{1\over 4}} (\sum_a X_a\,-
\sum_{\bar a} X_{\bar a})$.

    The solutions for $X_a$ and $X_{\bar b}$ can in principle each have two
values:
\be
X_a= \sigma_-\pm \sqrt{\sigma_-^2-\ft14 \lambda}\,,\qquad
X_{\bar a}= -\sigma_-\pm \sqrt{\sigma_-^2 -\ft14\lambda}\,.\label{sol}
\ee
However, it follows from (\ref{exta}) and (\ref{ext}) that
\be
4 (X_a + X_{\bar a}) + \lambda \, (X_a^{-1} + X_{\bar a}^{-1})=0
\ee
for any $a$ and $\bar a$, and so the positivity of the $X_a$ and
$X_{\bar a}$ implies that $\lambda<0$.  Consequently,
the positivity of $X_a$ and $X_{\bar a}$ implies that
the plus signs must be chosen for both equations in (\ref{sol}).  Thus
we must have all $X_a$ equal, $X_a\equiv X$, and all $X_{\bar a}$ equal,
$X_{\bar a} \equiv \bar X$, at any valid stationary point.  From Eqs.(\ref{sol})
it then follows we shall have
\be
\lambda = - 4 X\, \bar X\,,\qquad (p-2)\, X = (q-2)\, \bar X\,.
\ee
There is therefore a special case in which $q=2$ and hence also $p=2$;
otherwise, it must be that $q=2r\ge 4$, and $p\ge 3$.

   For $q=2r\ge4$, the explicit solution (\ref{sol})), subject to the
constraint (\ref{cons}), yields the result:
\be
X_a= X=\left( {{q-2}\over {p-2}} \right)
^{{q\over{p+q}}}\,,\qquad
X_{\bar a}= \bar X= \left(
{{p-2}\over {q-2}} \right) ^{{p\over{p+q}}}\,,
\ee
and the potential at the extremum has a positive value:
\be
\wtd V_0=2\, (p+q)\, \left({{q-2}\over
{p-2}}\right)^{{q-p}\over{q+p}}\,.
\ee

Note that the extremum of $\tilde V$ always corresponds to the positive value of
the potential. One can also prove that this extremum is always a {\it saddle
point} of the potential.  Note that this result is consistent with a  general
argument that non-compact reductions produce extrema of the scalar potential
that have always tachyonic direction \cite{Kallosh}.
(See however \cite{Fre} where the non-compact gauging
produced an example of de Sitter vacuum that is a minimum of the potential.)

   The special case when $p=q=2$ leads to $X\, \bar X=1$ and $\lambda =
-4$ at the stationary point.  The value of $X=\bar X^{-1}$ is undetermined,
meaning there is a ``flat direction,''
and the potential on this line of stationary points is given by
\be
\wtd V_0 = 8\,.
\ee

   Cases that arise in supergravities are associated with consistent
Pauli reductions on  spheres: $S^2$, $S^3$, $S^4$, $S^5$ or $S^7$ (or their
non-compact versions where $S^{p+q-1}$ is replaced by
${\cal H}^{p,q}$\ ), and thus with
\be
p+q = 3, 4, 5, 6, 8\,.
\ee
Given our findings above, namely that stationary points for $\wtd V$
arise in the non-compact cases for $q=2$ with $p=2$, or $q=2r\ge 4$
with $p\ge 3$, we see that
stationary points will arise only for two non-compact gauge groups
associated with consistent Pauli reductions on hyperboloidal spaces,
namely $SO(2,2)$ and $SO(4,4)$.  (Recall that we can always take $p\ge
q$.)  These are associated respectively with the replacements of the
following spheres by the corresponding hyperboloidal spaces:
\be
S^3 \longrightarrow {\cal H}^{2,2}\,,\qquad
S^7 \longrightarrow {\cal H}^{4,4}\,.\label{s3s7}
\ee
The first of these arises in the consistent Pauli reduction of type I
ten-dimensional supergravity to give $N=2$ gauged supergravity in
$D=7$; this reduction was first derived, for the compact choice $S^3$,
in \cite{cvluposata}.  Its non-compact analogue, with the consistent
reduction on ${\cal H}^{2,2}$, was recently studied in
\cite{cvegibpop}; the resulting $N=2$, $SO(2,2)$ gauged
seven-dimensional supergravity was used in order to obtain the
Salam-Sezgin \cite{salsez} 
$N=(1,0)$ gauged supergravity in six dimensions, by means
of a further $S^1$ reduction and consistent chiral truncation
\cite{cvegibpop}.

    The second case in (\ref{s3s7}) arises in the consistent $S^7$
reduction of eleven-dimensional supergravity \cite{dewitnic}.  The
investigation of the truncations that can be made by reducing instead
on ${\cal H}^{4,4}$ and then setting the scalar fields in $\wtd
T_{ij}$ to their  fixed values, corresponding to the extrema of the potential
$\tilde V$,  will  be studied in section
\ref{h44sec} below.

\subsection{Compact case}

   For completeness we shall also analyse the extrema of the
potential arising from the compact cases, such as those arising in the
case of consistent sphere reductions. These correspond to taking $q=0$ in
the discussion of section \ref{noncompsec}.  If a consistent reduction on
the sphere $S^{p-1}$ exists, it will give rise to a scalar potential with
an $SO(p)$ symmetry.  Again, we focus on the case where $\wtd T_{ij}$
is unimodular.  At the extrema of the potential $\tilde V$  will again
a truncation of the the scalar fields  ${\tilde T_{ij }}$  to the their fixed
valued at these extrema.

Stationary points of the potential (\ref{pot}) (with $q=0$) will be governed by (\ref{exta}), where
$\sigma_-\equiv \ft14 \sum X_a$.
   It follows from (\ref{exta}) that
\be
2(p-2)\, \sum_a X_a = \lambda \sum_a X_a^{-1}\,,
\ee
and therefore that $\lambda\ge 0$ (since in order to have a non-trivial
situation, we must certainly have $p\ge 2$).  It then  follows that the solutions
of (\ref{exta}), namely
\be
X_a = \sigma_- \pm \sqrt{\sigma_-^2 - \ft14\lambda}\,
\ee
can be positive for either choice of sign. Thus in principle we can have
\be
X_a= \a\,,\quad 1\le a\le m\,;\qquad X_a = \beta\,,\quad m+1 \le a \le m+n
\,,
\ee
where $\a\equiv \sigma_- + \sqrt{\sigma_-^2 - \ft14\lambda}$,
$\beta \equiv \sigma_- - \sqrt{\sigma_-^2 - \ft14\lambda}$ and   $m+n=p$.

   There are now two possible sets of solutions. The first corresponds
to taking all $X_a$ equal, in which case without loss of generality we
may take $n=0$ and so $X_a=\a$ for all $a$.  From the unimodularity of
$\wtd T_{ij}$ it then follows that $\a\equiv 1$ and hence we have
\be
X_a=1\,,\qquad \wtd V_0 = -p\, (p-2)\,.
\ee
The second possibility, in which unequal values $\a$ and $\beta$ occur
for non-vanishing numbers $m$ and $n=p-m$ of the $X_a$, implies that
\be
(m-2)\, \a + (n-2)\, \beta =0\,, \label{ab}
\ee
and hence positivity of the $X_a$'s (i.e. $\a>0$ and $\beta>0$)  implies
 that the only  remaining  solutions of (\ref{ab}) are those corresponding to
$m=1$  ($n=p-1\ge 3$) or $n=1$ ($m=p-1\ge 3$).  Choosing, without loss
of generality, $m=1$, we then
find
\be
X_1=\a = (p-3)^{\ft{p-1}{p}}\,,\quad X_a=\beta =(p-3)^{-\ft{1}{p}} \,,
\quad  2 \le a \le p\, ,\quad
\wtd V_0= -2p\, (p-3)^{\ft{p-2}{p}}\,.
\ee
Note that for $p<4$, the only extremum of the potential is the ``trivial''
one with all  $X_a=1$. On the other hand for $p\ge 4$, the potential has two
extrema with the property that the  ``trivial'' one   has always a less negative
cosmological constant.  In the context of the renormalization group flow
(associated with the dual field theory) the flows  start in the ultra-violet
regime at the trivial minimum and run toward the non-trivial one in the
infra-red regime.

\section{Pauli Reductions on Hyperboloidal Spaces ${\cal H}^{p,q}$}

   In this section, we shall enumerate some examples of supergravities
with non-compact gaugings that can be obtained by means of consistent
Pauli reductions on the hyperboloidal spaces ${\cal H}^{p,q}$.  These
examples are in one-to-one correspondence with already known cases of
supergravities with compact gaugings coming from consistent Pauli
sphere reductions.  The hyperboloidal reductions can in fact be
obtained by making analytic continuations of the existing sphere
reductions.  An equivalent, and rather more elegant approach, is first
to rewrite the sphere-reduction examples in a notation where the
passage from the compact to non-compact internal space is accomplished
merely by a replacement of a Euclidean-signature metric on the gauge
group by an indefinite-signature metric. As a consequence, the
non-linear Kaluza-Klein Ans\"atze, both in the compact and the
non-compact cases involve only real values for the p-form field
strengths, and positive definite metric in the internal space. In
addition the resulting lower dimensional theories contain only fields
with positive definite kinetic energy.

   A detailed enumeration of theories where a consistent Pauli sphere
reduction is known to exist was given in \cite{consistent}.  These are
examples of dimensional reductions on coset spaces, which in practice
are usually spheres.  If no fields were truncated out in the
dimensional reduction, the process of dimensional reduction would
necessarily be consistent, and one would end up with infinite towers
of massive fields as well as a finite number of massless fields that
included the metric and the gauge bosons of the isometry group of the
internal coset space.  Generically, it is inconsistent to set the
infinite towers of massive fields to zero, because non-linear terms
built from the massless fields that are retained will act as sources
for the massive fields that one wants to set to zero.  By a Pauli
reduction we mean a reduction in which one {\it can},
exceptionally, consistently set all the massive fields to zero, with
the set of lower-dimensional fields that are retained including the
Yang-Mills gauge bosons associated with the entire isometry group of
the internal coset manifold.  Thus for a Pauli reduction on the sphere
$S^n$, the retained lower-dimensional fields would include the
Yang-Mills gauge fields for the group $SO(n+1)$.  The success of a
consistent Pauli reduction depends on remarkable ``conspiracies'' between
properties of the internal coset space and properties of the theory
one is reducing.

   The list of consistent Pauli reductions presented in \cite{consistent}
comprised a number of examples with internal spaces $S^n$.  The fields
that are retained in the reduction include the metric, the gauge bosons
$A_\1^{ij}$ of $SO(n+1)$, and scalars described by a symmetric $(n+1)\times
(n+1)$ matrix $T_{ij}$:

\bigskip\bigskip
\centerline{
\begin{tabular}{|c|c|c|c|c|c|c|}\hline
$p$-form & Dilaton & Higher-Dim & Lower-Dim. & Sphere & Gauge
Group &Extra fields\\ \hline\hline $F_\2$ & Yes & Any $D$ & $D-2$
& $S^2$ & $SO(3)$ &None \\ \hline $F_\3$ & Yes & Any $D$ & $D-3$ &
$S^3$ & $SO(4)$ & $A_\2$\\ \hline $F_\3$ & Yes & Any $D$ & $3$ &
$S^{D-3}$ & $SO(D-2)$&None \\ \hline $F_\4$ & No & 11 & 7 & $S^4$
& $SO(5)$ &$A_\3^i$\\ \hline $F_\4$ & No & 11 & 4 & $S^7$ &
$SO(8)$ &$\phi_{[ijk\ell]_+}$\\ \hline $F_\5={*F_\5}$ & No & 10 &
5 & $S^5$ & $SO(6)$ &None\\ \hline
\end{tabular}}
\bigskip

\noindent{\bf Table 1:}  Consistent Pauli reductions on $S^n$,
retaining $SO(n+1)$ gauge fields.
The last column indicates what additional fields, beyond the
metric, the gauge fields $A_\1^{ij}$ and the scalars $T_{ij}$, are
massless, and must therefore be included, in a consistent
truncation.  The Table is taken from \cite{consistent}.
\bigskip

   The first row in Table 1 corresponds to Pauli reductions of
an Einstein-Maxwell-dilaton system in $D$ dimensions on $S^2$.  The
second and third rows correspond to Pauli reductions of the low-energy
effective theory of the bosonic string in $D$ dimensions on $S^3$ or
$S^{D-3}$.  The fourth and fifth rows correspond to the Pauli
reduction of eleven-dimensional supergravity on $S^4$ or $S^7$, and the last
row corresponds to the Pauli reduction of type IIB supergravity on
$S^5$.

    Each of these examples has its own particular features, and one
cannot give a ``universal'' Pauli-reduction ansatz that encapsulates
them all in a single set of formulae.  In particular, the set of
additional fields that might need to be retained in order to
achieve a consistent reduction is highly theory-specific.  For example,
in the $S^7$ reduction of eleven-dimensional supergravity one must
retain an additional massless 35 pseudo-scalar fields, as well as
the 35 scalars described by $T_{ij}$.  We shall not attempt, therefore
to present general formulae, nor shall we present the known details
in all the above cases.  Rather, we shall take one of the Pauli
reductions as an example, in order to show how the previously obtained
reduction formulae can be straightforwardly modified to generalise
from the compact case where the reduction is on $S^n$ to the
non-compact case where the reduction is on ${\cal H}^{p,q}$, where
$n=p+q-1$.  It should then be clear how the analogous transitions are
achieved in all the other examples.

   We shall take for our example the consistent $S^4$ Pauli reduction
of eleven-dimensional supergravity.  The complete reduction ansatz
was derived in \cite{vann}; it was re-expressed in a form that we
shall adopt here in \cite{cvluposata}.  The reduction ans\"atze for
the eleven-dimensional metric and 4-form filed strength are given by
\bea
d\hat s_{11}^2 &=& \Delta^{1/3}\, ds_{7}^2 + \fr1{g^2}\Delta^{-2/3}\,
T^{-1}_{AB}\, D\mu^A\, D\mu^B\,,\label{metel}
\eea
\bea
\hat F_\4 &=& \fft1{4!}\, \ep^{\phantom{\Sigma_\Sigma}}_{A_1\cdots A_5}\,
\Big[- \fr1{g^3} U\, \Delta^{-2} \mu^{A_1} D\mu^{A_2}\wedge \cdots \wedge
D\mu^{A_5}\nn\\
&& + \fr4{g^3} \Delta^{-2}\, T^{A_1}{}_B\, DT^{i_2}{}_C\,
\mu^B\, \mu^C\,
D\mu^{A_3}
\wedge \cdots \wedge D\mu^{A_5}\label{4form}\\
&& + \fr6{g^2} \Delta^{-1} F_\2^{A_1 A_2} \wedge
D\mu^{A_3}\wedge D\mu^{A_4}\, T^{A_5}{}_B\, \mu^jB \Big] - T_{AB}\,
{*S_\3^A}\, \mu^B + \fft1{g}\, S_{\3\, A} \wedge D\mu^A\,,\nn
\eea
where
\bea
U \equiv 2 T_{AC}\, T_{BD}\,\eta^{CD}\, \mu^A\, \mu^B
- \Delta\, T_{AB}\, \eta^{AB}\,, \qquad
\Delta \equiv T_{AB}\, \mu^A\, \mu^B\,,\nn\\
F_{\2\, A}{}^B \equiv dA_{\1\, A}{}^B + g\, A_{\1\, A}{}^C
\wedge A_{\1\, C}{}^B\,,
\qquad  D\mu^A \equiv d\mu^A + g\,  A_{\1}^A{}_B\, \mu^B\,,\nn\\
DT_{AB} \equiv dT_{AB} + g\, A_{\1\, A}{}^C\, T_{CB} +
g\, A_{\1 \, B}{}^C\, T_{AC}\,,
\qquad \mu^A\, \mu^B \, \eta_{AB} \equiv 1\,,
\eea
where the symmetric matrix $T_{AB}$, which parameterises the scalar
coset $SL(5,\R)/SO(5)$, is unimodular.

   Aside from a small change of notation, the only difference between
the ansatz above and the one presented in \cite{cvluposata} is that
in the latter the gauge group was taken to be $SO(5)$, meaning that
$\eta_{AB}=\delta_{AB}$, whereas here $\eta_{AB}$ is allowed to have
indefinite signature $(p,q)$, $p+q=5$, corresponding to an
$SO(p,q)$ gauging.  Note that all $A,B,\ldots$ gauge-group indices
are raised and lowered with $\eta_{AB}$.

   Substituting the above ansatz into the equations of motion
of eleven-dimensional supergravity, one finds that the lower-dimensional
fields satisfy the equations of motion of $SO(p,q)$-gauged $N=4$ supergravity
in seven dimensions, which follow from the Lagrangian
\bea
{\cal L}_7 &=& R\, {*\oneone} - {*P^{ij}}\wedge P^{ij}
-\ft1{4}\, T^{-1}_{AC}\, T^{-1}_{BD}\, {* F_\2^{AB}}\wedge F_\2^{CD}
-\ft12 T_{AB}\, {*S_\3^A}\wedge S_\3^B \nn\\
&&+ \fft1{2g} S_\3^A\wedge D S_\3^B \,\eta_{AB} -
\fft1{8g}  \ep^{\phantom{\Sigma_\Sigma}}_{A B_1\cdots B_4}\,
S_\3^A\wedge F_\2^{B_1 B_2}\wedge
F_\2^{B_3 B_4} + \fr1g \Omega_\7 - V\, {*\oneone}\,,\label{d7lag}
\eea
where
\be
P_{ij}\equiv \Pi^{-1}{}_{(i}{}^A\, (\delta_A{}^B\, d + g\, A_{\1 \, A}{}^B) \Pi_B{}^k
\,\delta_{j) k}
\ee
and the potential $V$ is given by
\be
V = \ft12  g^2 \Big(2 T_{ij}\, T_{ij} - (T_{ii})^2 \Big)\,.
\ee
Note that $T_{ij}$, with $SO(5)_c$ indices, and $T_{AB}$,
with $SO(p,q)_g$ indices, are given in terms of the scalar vielbein
$\Pi^i_A$ by
\be
T_{ij} = \Pi_i^{-1\, A}\, \Pi_j^{-1\, B}\, \eta_{AB}\,,\qquad
T^{AB} = \Pi^{-1}{}_i{}^A\, \Pi^{-1}{}_i{}^B\,.\label {ts}
\ee
The form of the Chern-Simons term $\Omega_\7$, built from the
Yang-Mills fields, can be found in \cite{cvluposata}.

    The geometry of the ``internal'' manifold can easily be seen from
the above expressions.  From the metric reduction ansatz in (\ref{metel}),
we see that if we consider the situation where the scalars and Yang-Mills
fields are taken to be trivial, meaning in particular that $T_{AB}=
\delta_{AB}$ (see (\ref{ts})), we shall have
\be
d\hat s_{11}^2 = \Delta^{1/3}\, ds_7^2 + \fft{1}{g^2}\, \Delta^{2/3}\,
\delta_{AB}\, d\mu^A\, d\mu^B\,,
\ee
where
\be
\Delta = \delta_{AB}\, \mu^A\, \mu^B\,,
\ee
and, of course, $\eta_{AB}\, \mu^A\, \mu^B=1$. Thus the internal
metric here is positive definite, and its isometry group is the
intersection of $SO(p+q)=SO(5)$, which leaves the Euclidean metric
$\delta_{AB}$ invariant, and $SO(p,q)$, which leaves $\eta_{AB}$ invariant.
This intersection is $SO(p)\times SO(q)$.

    We have presented the case of the consistent Pauli reductions of
eleven-dimensional to $D=7$ as an explicit example.  With appropriate
changes, one can straightforwardly discuss all the other known consistent
Pauli reductions.

One generic property of these reduction is the appearance of the
scalar field potential, which has a universal form of the type
(\ref{poten}).

\section{de Sitter-type Supergravities in Four Dimensions}\label{h44sec}

    In this section, we shall study explicit examples of four-dimensional
non-compact gauged supergravities,  which can be obtained
by a process of analytic continuation, and their associated embeddings in
eleven dimensions via consistent reductions.

\subsection{$N=4$ de Sitter gauged theory}

   For this construction, we shall begin with the $SO(4)$ gauged $N=4$
supergravity in four dimensions, and then perform an analytic
continuation to a de Sitter type supergravity.  By continuing the
known consistent $S^7$ reduction from $D=11$, we shall show how the
de Sitter supergravity arises as a consistent reduction on ${\cal H}^{4,4}$.

   To begin, let us consider the bosonic sector of the four-dimensional
$N=4$ gauged $SO(4)$ supergravity.  In the notation of \cite{d4gauge},
the bosonic Lagrangian may be written as
\bea
{\cal L}_4 &=& R\, {*\oneone} - \ft12 {*d\phi}\wedge d\phi - \ft12
e^{2\phi}\, {*d\chi}\wedge d\chi - V\, {*\oneone} \nn\\
&&- \ft12 e^{-\phi}\, {*F_\2^i}\wedge F_\2^i  -\ft12 \fft{e^\phi}{1+
\chi^2\, e^{2\phi}}\, {* \wtd F_\2^i}\wedge \wtd F_\2^i\,,\label{d4lag}\\
&&- \ft12\chi\, F_\2^i\wedge F_\2^i +
\ft12 \fft{\chi\, e^{2\phi}}{1+\chi^2\, e^{2\phi}}\, \wtd F_\2^i \wedge
\wtd F_\2^i\,,\nn
\eea
where the potential $V$ is
\be
V = -2g^2\, (4+ 2 \cosh\phi + \chi^2\, e^\phi)\,,\label{pot0}
\ee
and
\be
F_\2^i = dA_\1^i + \ft12 g\, \ep_{ijk}\, A_\1^j\wedge A_\1^k\,,\qquad
\wtd F_\2^i = d\wtd A_\1^i + \ft12 g\, \ep_{ijk}\, \wtd A_\1^j
\wedge \wtd A_\1^k\,.\label{fde0}
\ee

   We now perform the following continuations:
\be
g\longrightarrow -\im\, g\,,\quad A_\1^i\longrightarrow \im\, A_\1^i\,,\quad
\quad \wtd A_\1^i\longrightarrow \im\, \wtd A_\1^i\,,\quad
\phi\longrightarrow \phi + \im\, \pi\,.\label{cont1}
\ee
After doing this, the Lagrangian (\ref{d4lag}) retains the same
form
\bea
{\cal L}_4 &=& R\, {*\oneone} - \ft12 {*d\phi}\wedge d\phi - \ft12
e^{2\phi}\, {*d\chi}\wedge d\chi - V\, {*\oneone} \nn\\
&&- \ft12 e^{-\phi}\, {*F_\2^i}\wedge F_\2^i  -\ft12 \fft{e^\phi}{1+
\chi^2\, e^{2\phi}}\, {* \wtd F_\2^i}\wedge \wtd F_\2^i\,,\label{d4lag1}\\
&&- \ft12\chi\, F_\2^i\wedge F_\2^i +
\ft12 \fft{\chi\, e^{2\phi}}{1+\chi^2\, e^{2\phi}}\, \wtd F_\2^i \wedge
\wtd F_\2^i\,,\nn
\eea
except that now the potential $V$ in (\ref{pot0}) is replaced by
\be
V = 2g^2\, (4- 2 \cosh\phi - \chi^2\, e^\phi)\,,\label{pot1}
\ee
The $SU(2)\times SU(2)$ Yang-Mills fields are still given by the
same expressions (\ref{fde0}).
Note that the kinetic terms for all fields retain their conventional
signs, and thus the theory is still ghost-free. One can see very clearly
in
this example that it is the presence of the couplings of the scalar
fields in the Yang-Mills kinetic terms that allows their signs to
remain the standard ones despite the continuations $A_\1 \longrightarrow
\im\, A_\1$, owing to the compensating sign changes induced by
the continuation $\phi\longrightarrow \phi + \im\, \pi$.
However, the
scalar potential, which in the original compact form (\ref{pot0}) had
a minimum at $(\phi=0,\chi=0)$ with $V_0= -12g^2$, now has a minimum
at $(\phi=0,\chi=0)$ with $V_0 = + 4 g^2$.

   It should be noted that after the analytic continuations the gauge
group continues to be the compact group $SO(4) \sim SU(2)\times SU(2)$.
If we had performed an analogous analytic continuation on the full
$N=8$ gauged $SO(8)$ supergravity of de Wit and Nicolai \cite{dewitnic},
we would have obtained the non-compact gauging with $SO(4,4)$.  This
would be subject to a spontaneous symmetry breaking to its
compact $SO(4)\times SO(4)$ subgroup, and in fact the gauge fields that
are retained in the truncated $N=4$ theory that we are considering
here reside entirely within one of these $SO(4)$ factors, and hence
only a compact gauge group is seen here.

   The embedding of this de Sitter-type $N=4$ gauged supergravity in
$D=11$ can be seen by performing the corresponding analytic continuations
in the $S^7$ reduction formulae obtained in \cite{d4gauge}. From
the formulae in section 2 of \cite{d4gauge}, we see that after implementing
the continuations (\ref{cont1}) on the four-dimensional fields, we should
also make the continuation $\xi\longrightarrow \im\, \xi$ on the
``azimuthal'' coordinate of the description of $S^7$ as a foliation
of $S^3\times S^3$ surfaces.  This results in the consistent reduction
ansatz\footnote{Note that when performing the analytic continuation
of the expressions in \cite{d4gauge}, the sign of the entire
eleven-dimensional metric reverses, owing to a sign change of
the quantity $\Delta^2$ defined there.  This must be compensated by
using the ``trombone'' scaling symmetry of the eleven-dimensional
theory, under which $\hat g_{MN} \longrightarrow \lambda^2\, \hat g_{MN}$,
$\hat A_{MNP} \longrightarrow \lambda^3\, \hat A_{MNP}$, $\hat\psi_M
\longrightarrow \lambda\, \hat\psi_M$.  This is a
symmetry of the $D=11$ equations of motion, corresponding to a
homogeneous constant scaling of the action.  Specifically, we shall
take $\lambda=-\im$.  As we shall see below, the associated imaginary
rescaling of the antisymmetric tensor is precisely what is needed in
order to obtain a real expression after the continuations.}
\be
d\hat s_{11}^2 = \Delta^{\fft23}\, ds_4^2 + 2 g^{-2}\,
\Delta^{\fft23}\, d\xi^2
+ \ft12
g^{-2}\, \Delta^{\fft23}\, \Big[\fft{c^2}{c^2\, X^2 + s^2}\, \sum_i (h^i)^2
+ \fft{s^2}{s^2\, \tX^2 + c^2}\, \sum_i (\td h^i)^2\Big]\,,\label{metans}
\ee
where
\bea
&&X\equiv e^{\ft12\phi}\,,\qquad
\tX \equiv  X^{-1}\, q\,,\qquad q^2 \equiv 1 + \chi^2\,
X^4\,,\nn\\
&&\Delta \equiv  \Big[(c^2\, X^2 + s^2)(s^2\, \tX^2 + c^2)
\Big]^{\fft12} \,,\label{defs1}\\
&&c\equiv \cosh\xi\,,\qquad s\equiv \sinh\xi\,,\nn\\
&& h^i \equiv \sigma_i - g\, A_\1^i\,,\qquad \td h^i \equiv \td\sigma_i
-g\, \wtd A_\1^i\,.\nn
\eea
The three quantities $\sigma_i$ are left-invariant 1-forms on
$S^3=SU(2)$, and the three $\td\sigma_i$ are left-invariant 1-forms
on a second $S^3$.    They satisfy
\be
d\sigma_i = -\ft12 \ep_{ijk}\, \sigma_j\wedge \sigma_k\,,\qquad
d\td\sigma_i = -\ft12 \ep_{ijk}\, \td\sigma_j\wedge \td\sigma_k\,.
\label{leftinv}
\ee

    The reduction ansatz for $\hat F_\4$ given in \cite{d4gauge}
becomes
\be
\hat F_\4 = -g\, \sqrt2\, U\, \ep_\4 -
\fft{4s\, c}{g\,\sqrt2}\, X^{-1}\, {*dX}\wedge
d\xi + \fft{\sqrt2 s\, c}{g}\, \chi\, X^4\, {*d\chi}\wedge d\xi +
\hat F_\4' + \hat F_\4''\,,\label{fans1}
\ee
where
\be
U = -X^2\, c^2 + \tX^2\, s^2  + 2 \,,
\ee
and $\hat F_\4' = d\hat A_\3'$, with
\be
\hat A_\3' = f\, \ep_\3 + \td f\, \td\ep_3\,,\label{a3p}
\ee
where $\ep_\3 = \ft16 \ep_{ijk} \, h^i\wedge h^j\wedge h^k$ and
$\td\ep_\3 = \ft16 \ep_{ijk} \, \td h^i\wedge \td h^j\wedge \td h^k$.
The functions $f$ and $\td f$ are given by
\bea
f &=& \fft1{2\sqrt2}\, g^{-3}\, c^4 \chi\, X^2\, (c^2\, X^2+s^2)^{-1}\,,\nn\\
\td f &=& \fft1{2\sqrt2}\,
g^{-3}\, s^4\, \chi\, X^2\,  (s^2\, \tX^2 + c^2)^{-1}\,.
\label{ftf}
\eea
The terms in $\hat F_\4''$ comprise those involving the
$SU(2)\times SU(2)$ Yang-Mills field strengths $F_\2^i$ and $\wtd
F_\2^i$.  These are given by
\bea
\hat F_\4''&=& \ft1{\sqrt2}\, g^{-2}\, X^{-2}\, (-s\, c\, d\xi\wedge h^i
+\ft14 c^2\, \ep_{ijk}\, h^j\wedge h^k)\wedge (
{*F_\2^i} - \chi\, X^2\, F_\2^i)\nn\\
&&
+\ft1{\sqrt2}\, g^{-2}\, \tX^{-2}\, (s\, c\, d\xi\wedge \td h^i
-\ft14 s^2\, \ep_{ijk}\, \td h^j\wedge \td h^k)\wedge (
{*\wtd F_\2^i} + \chi\, X^2\, \wtd F_\2^i)\,.\label{fpp}
\eea
Note that in obtaining these real expressions for $\hat F_\4$, we made
use of the overall rescaling $\hat A_{MNP} \longrightarrow \im\,
\hat A_{MNP}$ that we discussed in the previous footnote.

    It is instructive to look at the nature of the internal 7-metric
in the ``ground state'' where $\phi=\chi=0=A_\1^i=\wtd A_\1^i$.   From
(\ref{metans}) we see that we shall have $d\hat s_{11}^2 = \Delta^{2/3}\,
ds_4^2 + 2\Delta^{-1/3}\, ds_7^2$ with $\Delta= \cosh2\xi$ and
\be
ds_7^2 =\cosh(2\xi)\, d\xi^2 + \ft14 \cosh^2\xi\, \sigma_i^2
      + \ft14 \sinh^2\xi\, \td\sigma_i^2\,.
\ee
This is precisely the standard ``undistorted'' positive-definite metric on the
the hyperboloid ${\cal H}^{4,4}$.  This can be seen by expressing the
coordinates $\mu^i=(\mu_a,\mu_{\bar a})$ on $\R^8$,
subject to the hyperboloidal constraint $\mu_a\, \mu_a -\mu_{\bar a}\,
\mu_{\bar a}=1$ as
\be
\mu_a = u_a\, \cosh\xi\,,\qquad \mu_{\bar a} = v_{\bar a}\,
\sinh\xi \,,
\ee
where $u_a\, u_a=1$ and $v_{\bar a}\, v_{\bar a}=1$ define two 3-spheres,
and substituting into the positive definite  metric $ds^2 = d\mu_a\,
d\mu_a
+ d\mu_{\bar a}\, d\mu_{\bar a}$ on $\R^8$.

     The theory that we have obtained here as a consistent reduction
is the bosonic sector of an 
$N=4$ de Sitter-type supergravity.  This can be consistently truncated to 
$N=3$, by setting the two sets of
$SU(2)$ gauge fields equal, and at the same time setting $\phi=\chi=0$.
In order to keep a canonical normalisation for the remaining $SU(2)$
Yang-Mills fields, we should also send $A_\1^i\longrightarrow
\ft1{\sqrt2}\,  A_\1^i$, $g\longrightarrow \sqrt2\, g$.
Upon doing so, we obtain the bosonic Lagrangian
\be
{\cal L}_4 = R\, {*\oneone} - \ft12 {*F_\2^i}\wedge F_\2^i - 8g^2\,
{*\oneone}\,,\label{n3lag}
\ee
where $F_\2^i = dA_\1^i + mg\, \ep_{ijk}\,A_\1^j\wedge A_\1^k$.
This bosonic sector of the truncated $N=3$ de Sitter supergravity 
is precisely the one that
was obtained recently in \cite{lupo}, together with its embedding in
eleven-dimensional supergravity.  In that work, the embedding of
the theory was derived from scratch.  It is interesting that by obtaining
the theory as the $N=3$ truncation of the larger $N=4$ theory, we can make
use of previous results in the literature \cite{d4gauge} in order to
establish the reduction procedure from $D=11$.  However if one truncates
to $N=3$ {\it before} looking at the embedding in $D=11$, the absence of
the scalar fields precludes one from implementing the analytic
continuation in (\ref{cont1}) that allowed us to perform a continuation
of the $N=4$ $S^7$ reduction of \cite{d4gauge}.

   Further truncations to lesser supersymmetry are also possible. One
can, for example, consider a truncation to $N=1$, in which one retains
just a Maxwell multiplet as well as the supergravity multiplet. In 
the bosonic sector, the Lagrangian is obtained from (\ref{n3lag}) by
retaining just a $U(1)$ gauge field, and so one has
\be
{\cal L}_4 = R\, {*\oneone} - \ft12 {*F_\2}\wedge F_\2 - 8g^2\,
{*\oneone}\,,\label{n1lag}
\ee
In fact it is presumably the case that this is the bosonic sector of
the axially-gauged $N=1$ de Sitter supergravity constructed in
\cite{freedman}.  (Since we have not explicitly studied the fermionic
sector here this remains conjectural at this stage, but the existence
of the non-compact gauged $N=8$ supergravities discussed in
\cite{Hull,hullwarn}, and of the axially-gauged $N=1$ supergravity
obtained in \cite{freedman}, lend credence to the conjecture.)  In
particular, this implies that any solution of four-dimensional
Einstein-Maxwell gravity with a positive cosmological constant can be
embedded in the de Sitter supergravity of \cite{freedman}, and hence,
via our consistent reduction, it can be lifted to a solution in
eleven-dimensional supergravity.  Examples of such four-dimensional
solutions include the cosmological multi back hole solutions of
\cite{KastorTraschen}.

\subsection{$N=2$ de Sitter gauged theories}\label{n2d4gaugesec}

   In this section, we shall consider $N=2$ de Sitter supergravities
obtained by starting with the four-dimensional $N=2$ supergravity
with $U(1)^4$ gauging, whose consistent embedding in $D=11$ supergravity
was discussed in \cite{spheresa}.   For simplicity, we shall follow
\cite{spheresa} and omit the three axionic scalar fields that form
part of the supergravity theory.  Their inclusion in the four-dimensional
theory itself is straightforward, and we refer the reader to Appendix B
of \cite{spheresa} for a discussion of the details.  Including the axions
in the consistent reduction from $D=11$ is a more difficult problem,
and we shall not attempt to address that here.

   With the axions omitted, the bosonic sector of the four-dimensional
Lagrangian for the $U(1)^4$ gauged theory is given by
\be
e^{-1}\, {\cal L}= R -\ft12(\del\vec\varphi)^2 - \ft14
\sum_{i=1}^4 X_i^{-2}\, (F^i)^2 -V\,,\label{u14lag0}
\ee
where $\vec\varphi=(\varphi_1,\varphi_2,\varphi_3)$, the scalar
potential is given by
\bea
V &=& - 4 g^2\, \sum_{i<j} X_i\, X_j\\
&=& - 8 g^2\, (\cosh\varphi_1 + \cosh\varphi_2 + \cosh\varphi_3)
  \,,\label{u14v0}\,\nn
\eea
and
\bea
&& X_1 = e^{-\ft12(\varphi_1 + \varphi_2 + \varphi_3)}\,,\qquad
X_2 = e^{-\ft12(\varphi_1 - \varphi_2 - \varphi_3)}\,,\nn\\
&& X_3 = e^{-\ft12(-\varphi_1 + \varphi_2 - \varphi_3)}\,,\qquad
X_4 = e^{-\ft12(-\varphi_1 - \varphi_2 + \varphi_3)}\,.
\eea

   The embedding of the $U(1)^4$ gauged theory in $D=11$ supergravity
was constructed in \cite{spheresa}; it involves a consistent Pauli-type
reduction on $S^7$, and is given by
\bea
d\hat s_{11}^2 &=& \Delta^{2/3}\, ds_4^2 + g^{-2}\, \Delta^{-1/3}\,
\sum_{i=1}^4 X_i^{-1}\, \Big(d\mu_i^2 + \mu_i^2\, (d\phi_i + g\, A^i)^2\Big)
\,,\label{metansu14}\\
\hat F_\4  &=& \sum_{i=1}^4 \Big(
2g\, (X_i^2\, \mu_i^2 - \Delta\, X_i)\, \ep_\4
+ \fft1{2g}\,X_i^{-1}\, {* dX_i}\wedge d(i\mu_i^2)i\nn\\
&&\qquad  - \fft1{2g^2}\,
X_i^{-2}\, d(\mu_i^2)\wedge (d\phi_i + g\, A^i)\wedge {*F^i}\Big)\,,
\label{f4ansu14}
\eea
where
\be
\Delta= \sum_{i=1}^4 X_i\, \mu_i^2\,,\qquad
\sum_{i=1}^4 \mu_i^2=1\,,\label{u14delta}
\ee
$\ep_\4$ is the volume form in the four-dimensional metric $ds_4^2$, and
$*$ denotes Hodge dualisation in the four-dimensional metric.
Note that the round 7-sphere arises when the scalars are trivial ($X_i=1$),
and is described in terms of the four constrained coordinates $\mu_i$ and
the four azimuthal angles $\phi_i$ by
\be
d\Omega_7^2 = \sum_{i=1}^4 ( d\mu_i^2 + \mu_i^2\, d\phi_i^2)\,.
\ee

   We shall now describe two inequivalent analytic continuations, one
of which corresponds to a replacement of $S^7$ by ${\cal H}^{4,4}$,
and the other to a replacement of $S^7$ by ${\cal H}^{6,2}$.  The
consistent reductions in these cases, obtained by appropriate analytic
continuations of the complete $S^7$ reduction of de Wit and Nicolai
\cite{dewitnic}, would yield $N=8$ supergravities with the non-compact
gaugings $SO(4,4)$ and $SO(6,2)$ respectively.  In our case, where we
start with the restricted $N=2$ gauged theory of supergravity coupled
to three vector multiplets, we are retaining only the $U(1)^4$ gauge
field in the Cartan subgroup of $SO(8)$.  After the analytic
continuations, in each case we will still have $U(1)^4$ gauge fields;
these are in the compact Cartan subgroups of $SO(4,4)$ and $SO(6,2)$
respectively.  We shall therefore refer to the two analytically
continued theories as the $SO(4,4)$ and $SO(6,2)$ cases respectively,
even though our truncations retain only $U(1)^4$ gauge fields.

    In accordance with our general results in section
\ref{extremasec}, the former theory will have a scalar potential with
a stationary point, whilst the latter will not.

\subsubsection{The $SO(4,4)$ case}

   To perform the analytic continuations in this case, we take
\be
\varphi_1 \longrightarrow \varphi_1 + \im\, \pi\,,\qquad
A^i \longrightarrow \im\, A^i\,,\qquad
g\longrightarrow - \im\, g\,,\label{cont2}
\ee
with $\varphi_2$ and $\varphi_3$ left unchanged.  This implies that we
shall have
\be
X_1\longrightarrow -\im\, X_1\,,\quad
X_2\longrightarrow -\im\, X_2\,,\quad
X_3\longrightarrow \im\, X_3\,,\quad
X_4\longrightarrow \im\, X_4\,.\label{cont3}
\ee
The Lagrangian (\ref{u14lag0}) will therefore retain the identical form,
except that now the potential (\ref{u14v0}) will be replaced by
\be
V = 8g^2\, (\cosh\varphi_2 + \cosh\varphi_3 - \cosh\varphi_1)\,.
\ee

   Turning now to the embedding in eleven-dimensional supergravity, we
make the corresponding continuations
\be
\mu_3\longrightarrow -\im\, \mu_3\,,\qquad
\mu_4\longrightarrow -\im\, \mu_4\,,
\ee
while leaving $\mu_1$ and $\mu_2$ unchanged.  This implies that we shall
have $\Delta\longrightarrow -\im\, \Delta$, for which we may define the
cube root so that
\be
\Delta^{1/3} \longrightarrow \im\, \Delta^{1/3}\,.
\ee
Finally, we perform a ``trombone'' rescaling $d\hat s_{11}^2
\longrightarrow \lambda^2\, d\hat s_{11}^2$, $\hat A_\3\longrightarrow
\lambda^3\, \hat A_\3$ with $\lambda=\im$.  We therefore arrive at the
metric and field-strength reductions
\bea
d\hat s_{11}^2 &=& \Delta^{2/3}\, ds_4^2 + g^{-2}\, \Delta^{-1/3}\,
\sum_{i=1}^4 X_i^{-1}\, \Big(d\mu_i^2 + \mu_i^2\, (d\phi_i + g\, A^i)^2\Big)
\,,\label{metans44}\\
\hat F_\4  &=& \sum_{i=1}^4 \eta_i\, \Big(
2g\, (X_i^2\, \mu_i^2 - \Delta\, X_i)\, \ep_\4
+ \fft1{2g}\,X_i^{-1}\, {* dX_i}\wedge d(i\mu_i^2)i\nn\\
&&\qquad  - \fft1{2g^2}\,
X_i^{-2}\, d(\mu_i^2)\wedge (d\phi_i + g\, A^i)\wedge {*F^i}\Big)\,,
\label{f4ans44}
\eea
where
\bea
\Delta &=& \sum_{i=1}^4 X_i\, \mu_i^2\,,\qquad
\mu_1^2 + \mu_2^2 - \mu_3^2 - \mu_4^2=1\,,\nn\\
\eta_i &=& (1,1,-1,-1)\,.
\label{r44}
\eea
Note that if the scalar fields are taken to be trivial ($X_i=1$),
the internal space has the positive-definite metric on ${\cal H}^{4,4}$
given by
\be
ds_7^2 = \sum_{i=1}^4 ( d\mu_i^2 + \mu_i^2\, d\phi_i^2)\,,
\ee
where the $\mu_i$ coordinates are subject to the constraint given in
(\ref{r44}).

\subsubsection{The $SO(6,2)$ case}\label{s062casesec}

   Our analytic continuation in this case is taken to be
\bea
&&\varphi_1 \longrightarrow \varphi_1 + \fft{\im\, \pi}{2}\,,\qquad
\varphi_2 \longrightarrow \varphi_2 + \fft{\im\, \pi}{2}\,,\qquad
\varphi_3 \longrightarrow \varphi_3 - \fft{\im\, \pi}{2}\,,\nn\\
&&A^i \longrightarrow e^{-\ft{\im}{4} \pi}\, A^i\qquad
g  \longrightarrow e^{\ft{\im}{4} \pi}\, g\,,
\eea
which implies
\be
X_1 \longrightarrow e^{-\ft{\im}{4} \pi}\, X_1\,,\quad
X_2 \longrightarrow e^{-\ft{\im}{4} \pi}\, X_2\,,\quad
X_3 \longrightarrow e^{-\ft{\im}{4} \pi}\, X_3\,,\quad
X_4 \longrightarrow e^{\ft{3\im}{4} \pi}\, X_4\,.
\ee
This leaves the form of the Lagrangian (\ref{u14lag0}) unchanged
except that now the scalar potential (\ref{u14v0}) is replaced by
\be
V = 8 g^2\, (\sinh\varphi_1 + \sinh\varphi_2 - \sinh\varphi_3)\,.
\label{so62v}
\ee
This has no stationary points, and it is unbounded from above and below.

   For the embedding in eleven-dimensional supergravity, we make the
corresponding continuation
\be
\mu_4 \longrightarrow e^{-\ft{\im}{2} \pi}\, \mu_4\,,
\ee
while leaving $\mu_1$, $\mu_2$ and $\mu_3$ unchanged.
This means that $\Delta$  defined in
(\ref{u14delta}) will be changed according to $\Delta \longrightarrow
\exp(-\ft{\im}{4}\pi)\, \Delta$, for which we shall have the
replacement in the cube root:
\be
\Delta^{1/3} \longrightarrow e^{-\ft{3\im}{4} \pi}\, \Delta^{1/3}\,.
\ee
Finally, we perform a trombone rescaling $d\hat s_{11}^2 \longrightarrow
\lambda^2\, d\hat s_{11}^2$, $\hat A_\3\longrightarrow
\lambda^3\, \hat A_\3$ of the eleven dimensional fields with $\lambda=
\exp(-\ft{\im}{4} \pi)$.  This leads to the following expressions for
the metric and field strength reductions:
\bea
d\hat s_{11}^2 &=& \Delta^{2/3}\, ds_4^2 + g^{-2}\, \Delta^{-1/3}\,
\sum_{i=1}^4 X_i^{-1}\, \Big(d\mu_i^2 + \mu_i^2\, (d\phi_i + g\, A^i)^2\Big)
\,,\label{metans62}\\
\hat F_\4  &=& \sum_{i=1}^4 \eta_i\, \Big(
-2g\, (X_i^2\, \mu_i^2 - \Delta\, X_i)\, \ep_\4
- \fft1{2g}\,X_i^{-1}\, {* dX_i}\wedge d(\mu_i^2)\nn\\
&&\qquad  + \fft1{2g^2}\,
X_i^{-2}\, d(\mu_i^2)\wedge (d\phi_i + g\, A^i)\wedge {*F^i}\Big)\,,
\label{f4ans62}
\eea
where
\bea
\Delta &=& \sum_{i=1}^4 X_i\, \mu_i^2\,,\qquad
\mu_1^2 + \mu_2^2 + \mu_3^2 - \mu_4^2=1\,,\nn\\
\eta_i &=& (1,1,1,-1)\,.
\label{r62}
\eea
Note that if the scalar fields are taken to be trivial ($X_i=1$),
the internal space has the positive-definite metric on ${\cal H}^{6,2}$
given by
\be
ds_7^2 = \sum_{i=1}^4 ( d\mu_i^2 + \mu_i^2\, d\phi_i^2)\,,
\ee
where the $\mu_i$ coordinates are subject to the constraint given in
(\ref{r62}).  Note also that, in accordance with our discussion of
the existence of stationary points in section \ref{extremasec}, the
scalar potential given in (\ref{so62v}) for this ${\cal H}^{6,2}$
reduction has no extrema.

\section{Non-compact Gauged Supergravities in Higher Dimension}

   In this section we discuss two further examples associated with
non-compact gauged supergravities, in five and seven dimensions.  Our
starting points are the $N=2$ gauged theories whose consistent
reductions from type IIB on $S^5$ and $D=11$ supergravity on $S^4$
respectively were discussed in \cite{spheresa}.  The five-dimensional
theory has $U(1)^3$ gauge fields in the Cartan subgroup of the $SO(6)$
isometry of $S^5$, while the seven-dimensional theory has $U(1)^2$
gauge fields in the Cartan subgroup of the $SO(5)$ isometry of $S^4$.

\subsection{Five-dimensional $N=2$ gauged supergravity}

   In \cite{spheresa}, the consistent Pauli reduction that yields
$N=2$ gauged supergravity coupled to two vector multiplets was given.  The
gauge fields lie in the $U(1)^3$ Cartan subgroup of the full $SO(6)$ gauge
group of the $N=8$ theory.  The Lagrangian for the bosonic sector of the
$N=2$ theory is given by
\be
e^{-1}\, {\cal L}_5 = R - \ft12(\del\varphi_1)^2 -\ft12(\del\varphi_2)^2
- \ft14 \sum_i X_i^{-2}\, (F^i)^2  +\ft14
\ep^{\mu\nu\rho\sigma\lambda}\, F^1_{\mu\nu}\, F^2_{\rho\sigma}\,
A^3_\lambda -V\,,
\label{d5lag}
\ee
where the scalar potential is given by
\be
V= -4 g^2 \, \sum_{i=1}^3 X_i^{-1}\,,\label{d5pot}
\ee
and we define
\be
X_1 = e^{-\ft1{\sqrt6}\varphi_1 -\ft1{\sqrt2}\varphi_2}\,,\qquad
X_2 = e^{-\ft1{\sqrt6}\varphi_1 +\ft1{\sqrt2}\varphi_2}\,,\qquad
X_3 = e^{\ft{2}{\sqrt6}\varphi_1}\,.
\ee

   The embedding in the type IIB ten-dimensional theory involves a
consistent Pauli reduction on $S^5$, described by \cite{spheresa}
\bea
d\hat s_{10}^2 &=& \Delta^{1/2}\, ds_5^2 + \fft1{g^2}\,
\Delta^{-1/2}\,  \sum_{i=1}^3 X_i^{-1}\,
\Big(d\mu_i^2 + \mu_i^2\, (d\phi_i +g\, A^i)^2\Big)\,,\nn\\
\hat G_\5 &=&\sum_{i=1}^3\Big( 2g(X_i^2\, \mu_i^2 - \Delta\, X_i)\,
\ep_\5 - \fft1{2g}\, X_i^{-1}\, {*dX_i}\wedge
d(\mu_i^2) \nn\\
&&+ \fft1{2 g^2} \,  X_i^{-2}\, d(\mu_i^2)\wedge (d\phi_i +g\, A_\1^i)
\wedge { * F_\2^i}\Big)\,,
\eea
where the self-dual 5-form is given by $\hat F_\5 = \hat G_5 +
{\hat * \hat G_5}$, and
\be
\Delta = \sum_{i=1}^3 X_i\, \mu_i^2\,,\qquad \sum_{i=1}^3\mu_i^2=1\,.
\ee

    One can perform an analytic continuation of the full $S^5$
reduction to describe an $N=8$  non-compact $SO(4,2)$ gauged supergravity.
In the $N=2$ truncation considered here, the retained $U(1)^3$ gauge
fields reside in the Cartan subgroup.  The appropriate
analytic continuation is achieved by sending
\be
\varphi_1 \longrightarrow \varphi_1 - \ft{2\im}{\sqrt6}\, \pi\,,\qquad
A^i\longrightarrow e^{\ft{\im}{3}\pi}\, A^i\,,\qquad
g\longrightarrow e^{-\ft{\im}{3}\pi}\, g\,,
\ee
with $\varphi_2$ unchanged.  Under these continuations, we shall have
\be
X_1 \longrightarrow e^{\ft{\im}{3}\pi}\, X_1\,,\qquad
X_2 \longrightarrow e^{\ft{\im}{3}\pi}\, X_2\,,\qquad
X_3 \longrightarrow e^{-\ft{2\im}{3}\pi}\, X_3\,.
\ee
The Lagrangian (\ref{d5lag}) becomes
\be
e^{-1}\, {\cal L}_5 = R - \ft12(\del\varphi_1)^2 -\ft12(\del\varphi_2)^2
- \ft14 \sum_i X_i^{-2}\, (F^i)^2  -\ft14
\ep^{\mu\nu\rho\sigma\lambda}\, F^1_{\mu\nu}\, F^2_{\rho\sigma}\,
A^3_\lambda -V\,,
\label{d5lag2}
\ee
(\ie the sign of the Chern-Simons term is reversed)
with the scalar potential (\ref{d5pot}) being replaced by
\bea
V&=& 4g^2 \, (X_1^{-1} + X_2^{-1} - X_3^{-1})\,,\nn\\
&=& 4g^2\, (2 e^{\ft1{\sqrt6}\varphi_1}\, \cosh \ft1{\sqrt2}\varphi_2 -
e^{-\ft2{\sqrt2}\varphi_1})\,.
\eea
It is easily seen that indeed, as expected from the discussion in
section (\ref{extremasec}), this potential has no stationary points.
It is unbounded from above and below.

   In the description of the embedding in the type IIB theory, we must
make corresponding continuation
\be
\mu_3 \longrightarrow e^{\ft{\im}{2}\pi}\, \mu_3\,,
\ee
leaving $\mu_1$ and $\mu_2$ unchanged, implying that we shall have
\be
\Delta\longrightarrow e^{\ft{\im}{3}\pi}\, \Delta\,.
\ee

   The type IIB theory has a ``trombone'' symmetry under which we
perform the rescalings $d\hat s_{10}^2 \longrightarrow \lambda^2\,
d\hat s_{10}^2$, and $\hat F_\5 \longrightarrow \lambda^4\, \hat F_\5$.
If we take $\lambda=\exp(-\ft{\im}{12}\pi)$, the we finally arrive at the
reduction
\bea
d\hat s_{10}^2 &=& \Delta^{1/2}\, ds_5^2 + \fft1{g^2}\,
\Delta^{-1/2}\,  \sum_{i=1}^3 X_i^{-1}\,
\Big(d\mu_i^2 + \mu_i^2\, (d\phi_i +g\, A^i)^2\Big)\,,\nn\\
\hat G_\5 &=&\sum_{i=1}^3\eta_i\, \Big( 2g(X_i^2\, \mu_i^2 - \Delta\, X_i)\,
\ep_\5 - \fft1{2g}\, X_i^{-1}\, {*dX_i}\wedge
d(\mu_i^2) \nn\\
&& +\fft1{2 g^2} \,  X_i^{-2}\, d(\mu_i^2)\wedge (d\phi_i +g\, A_\1^i)
\wedge { * F_\2^i}\Big)\,,
\eea
where
\bea
\Delta &=& \sum_{i=1}^3 X_i\, \mu_i^2\,,\qquad
\mu_1^2 + \mu_2^2 - \mu_3^2 =1\,,\nn\\
\eta_i &=& (1,1, -1)\,.
\eea
This describes a reduction on the non-compact hyperboloid ${\cal H}^{4,2}$,
and the supergravity we have obtained here is the $N=2$ truncation of
the $N=8$ non-compact gauged $SO(4,2)$ supergravity in five dimensions.

\subsection{Seven-dimensional $N=2$ gauged supergravity}

   The expression for a consistent $S^4$ reduction that yields
the $U(1)^2$ gauged $N=2$ supergravity was given in \cite{spheresa}.  The
gauge fields lie in the Cartan subgroup of $SO(5)$.  The bosonic
Lagrangian is given by
\be
e^{-1}{\cal L}_7 = R -\ft12 (\del\vec\varphi)^2  -
\ft14 \sum_{i=1}^2 X_i^{-2} \, (F_\2^i)^2 -V\,,
\label{d7lag0}
\ee
where the scalar potential $V$ is given by
\be V=-g^2\left( 4 X_1 X_2 + 2X_1^{-1}\, X_2^{-2} + 2 X_2^{-1}\, X_1^{-2}
+\ft12 -(X_1 X_2)^{-4}\right)\,,\label{d7pot} \ee
and
\be
X_1 = e^{-\ft1{\sqrt2}\varphi_1 + \ft{1}{\sqrt{10}}\varphi_2}\,,\qquad
X_2 = e^{\ft1{\sqrt2}\varphi_1 + \ft{1}{\sqrt{10}}\varphi_2}\,.
\ee

   The theory is obtained via a consistent Pauli reduction on $S^4$
\cite{spheresa}, with
\bea
d\hat s_{11}^2 &=& \wtd\Delta^{1/3}\, ds_7^2 + g^{-2}\, \wtd\Delta^{-2/3}\,
\Big(X_0^{-1}\, d\mu_0^2 + \sum_{i=1}^2 X_i^{-1}\, (d\mu_i^2 + \mu_i^2\,
(d\phi_i + g\, A_\1^i)^2) \Big)\ ,\label{s4metred}\\
{\hat *\hat F_\4} &=&
2g\,\sum_{\a=0}^2 \Big(X_\a^2\, \mu_\a^2 - \wtd\Delta\, X_\a \Big)\,
\ep_\7 + g\, \wtd\Delta\, X_0\, \ep_\7
+\fft1{2g}\, \sum_{\a=0}^2 X_\a^{-1}\, { * dX_\a}
\wedge d(\mu_\a^2) \nn\\
&&+\fft1{2g^2}\, \sum_{i=1}^2 X_i^{-2}\, d(\mu_i^2)\wedge
(d\phi_i + g\, A^i_\1) \wedge
{ * F_\2^i}\ ,\label{s4f4red}
\eea
where the auxiliary variable $X_0$ is defined by
$X_0\equiv (X_1 X_2)^{-2}$, and we have
\be
\Delta = \sum_{\a=0}^2 X_\a\, \mu_\a^2\,,\qquad
\sum_{\a=0}^2 \mu_\a^2  =1\,.
\ee

   An analytic continuation to a de Sitter like supergravity is given by
sending
\be
\varphi_1\longrightarrow \varphi_1 + \ft{\im}{\sqrt2}\pi\,,\qquad
\varphi_2\longrightarrow \varphi_2 + \ft{\im}{\sqrt{10}}\pi\,,\qquad
A^1\longrightarrow e^{\ft{3\im}{5}\pi}\, A^i\,,\qquad
g\longrightarrow e^{-\ft{3\im}{5}\pi}\, g\,.
\ee
These imply that we shall have
\be
X_0 \longrightarrow e^{-\ft{2\im}{5}\pi}\, X_0\,,\qquad
X_1 \longrightarrow e^{-\ft{2\im}{5}\pi}\, X_1\,,\qquad
X_2 \longrightarrow e^{\ft{3\im}{5}\pi}\, X_2\,.
\ee
Under this continuation, the Lagrangian is still given by (\ref{d7lag}),
except that now the scalar potential (\ref{d7pot}) is replaced by
\bea
V &=& g^2\, ( 4 X_1\, X_2 - 2 X_1^{-1}\, X_2^{-2} + 2 X_1^{-2}\,
X_2^{-1} + \ft12 (X_1\, X_2)^{-2})\,,\nn\\
&=& g^2\, (4 e^{-\ft3{\sqrt{10}}\varphi_2}\, \sinh\ft1{\sqrt2}\varphi_1
+ 4 e^{\ft2{\sqrt{10}}\varphi_2} + \ft12 e^{-\ft{4}{\sqrt{10}}\varphi_2}
)\,.
\eea
One can easily see that, as expected, this potential has no stationary
points. It is unbounded from above and below.

    The continuation of the embedding in eleven-dimensional supergravity
is obtained by making the corresponding continuation,
\be
\mu_0\longrightarrow \mu_0\,,\qquad
\mu_1\longrightarrow \mu_1\,,\qquad
\mu_2\longrightarrow e^{-\ft{\im}{2}\pi}\, \mu_2\,.
\ee
This implies that we will have $\Delta\longrightarrow
\exp(-\ft{2\im}{5}\pi)\, \Delta$.  Using the trombone rescaling symmetry
$d\hat s_{11}^2 \longrightarrow \lambda^2\, d\hat s_{11}^2$,
${\hat *\hat F_\4}\longrightarrow \lambda^6\, {\hat * \hat F_\4}$, with
$\lambda = \exp(\ft{\im}{15}\pi)$, we find that the reduction
(\ref{s4metred}), (\ref{s4f4red}) becomes
\bea
d\hat s_{11}^2 &=& \Delta^{1/3}\, ds_7^2 + g^{-2}\, \Delta^{-2/3}\,
\Big(X_0^{-1}\, d\mu_0^2 + \sum_{i=1}^2 X_i^{-1}\, (d\mu_i^2 + \mu_i^2\,
(d\phi_i + g\, A_\1^i)^2) \Big)\ ,\label{s4metredso32}\\
{\hat *\hat F_\4} &=&\!\!
 -2g\,\sum_{\a=0}^2 \eta_\a\, X_\a^2\, \mu_\a^2\, \ep_\7  + g\,
\Delta\,(X_0 + 2X_1 - 2X_2)\,
\ep_\7
- \! \fft1{2g} \sum_{\a=0}^2 \eta_\a\, X_\a^{-1}\, { * dX_\a}
\wedge d(\mu_\a^2) \nn\\
&& -\fft1{2g^2}\, \sum_{i=1}^2 \eta_i\, X_i^{-2}\, d(\mu_i^2)\wedge
(d\phi_i + g\, A^i_\1) \wedge
{ * F_\2^i}\,,\label{s4f4redso32}
\eea
where
\bea
\Delta &=&\sum_{\a=0}^2 X_\a\, \mu_\a^2\,,\qquad \mu_0^2 +
      \mu_1^2 - \mu_2^2=1\,,\nn\\
\eta_\a &=& (1,1,-1)\,.
\eea
The internal metric lives on the hyperboloidal space ${\cal H}^{3,2}$.
In a full consistent reduction on this space, obtained by analytically
continuing the consistent $S^4$ reduction \cite{vann} that gives the
$N=4$ gauged $SO(5)$ theory in seven dimensions, one would obtain the
$N=4$ non-compact $SO(3,2)$ gauged supergravity.  Our $N=2$ truncation
retains just the $U(1)^2$ gauge field in the Cartan subgroup of
$SO(3,2)$.

\section{Black Hole and Cosmological de Sitter Solutions}

    In this section, we shall derive charged black hole and cosmological
solutions for specific supergravities with non-compact gaugings in
$D=4$, 5 and 7.  In the examples considered, only the charges residing in
the Abelian subgroup of $SO(p, 2r)$ are turned on, yielding solutions
that can all be described within an $N=2$ truncation.  Specifically
the equations of motion for such multiply-charged solutions can be
solved in the case of the $N=2$ truncation of these supergravities in
$D=4$, $D=5$ and $D=7$.  These solutions are related to
the AdS charged black hole solutions of the $N=2$ truncations of the
respective $SO(8)$, $SO(6)$ and $SO(5)$ gauged supergravities.

 \subsection{Three-charge solutions of  $D=5$, $N=2$ gauged 
$SO(2,4)$ and $SO(4,2)$ supergravities}

    These solutions are closely related to the AdS black hole
solutions of the $N=2$ truncation of five-dimensional $SO(6)$ gauged
supergravity, coupled to the three Abelian vector supermultiplets. The
latter solutions were derived in \cite{becvsa}, and are of the form
\bea
ds_5^2 &=& -(H_1H_2H_3)^{-2/3}\, f\, dt^2 +
(H_1H_2H_3)^{1/3}\, (f^{-1}\, dr^2 + r^2 d\Omega_{3,k}^2)\ ,\nn\\
X_i&=& H_i^{-1}\, (H_1H_2H_3)^{1/3}\ ,\qquad
A^i_\1 = \sqrt{k}\, (1-H_i^{-1})\, \coth(\sqrt{k}\, \beta_i)\, dt\,,
\label{d3adsbh}
\eea
where
\be
f=k-\fft{\mu}{r^2} + 4g^2\, r^2\, (H_1H_2H_3)\, , \label{d3adsbh1}
\ee
and the harmonic functions $H_i$ are given by
\be 
H_i = 1 + \fft{\mu\, \sinh^2(\sqrt{k}\, \beta_i)}{k\, r^2}\ .\label{d3adsbh2}
\ee
Here, $k$ can be 1, 0 or $-1$, corresponding to the cases where the
foliations in the transverse space have the metric $d\Omega_{3,k}^2$
on the unit $S^3$, $T^3$ or ${\bf H}^3$, where ${\bf H}^3$ is the unit
hyperbolic 3-space of constant negative curvature. Note that in order
to satisfy the Einstein equations of motion, the constants $c_i$ in
the harmonic functions $H_i=c_i + O(1/r^2)$ in \cite{becvsa} were
taken to be $1$.  This ensured that the contribution from the scalars
$X_i$ to the right-hand side of the Einstein equations was compatible
with the metric contribution in the Einstein equations (see section
3.2 of \cite{becvsa} for details).

   For the $N=2$ truncation of the $SO(2,4)$ and $SO(4,2)$
supergravities with the potential of the form (\ref{d5pot}), one can
apply the analysis of \cite{becvsa} in a straightforward way. The
equations of motion are solved with the same ansatz
(\ref{d3adsbh})-(\ref{d3adsbh1}) for the metric, scalars and gauge
fields, except that now the harmonic functions take the form
\be
H_i = \eta_i + \fft{\mu\, \sinh^2(\sqrt{k}\, \beta_i)}{k\, r^2}\, , \qquad
i=1,\, 2,\, 3\, , 
\label{d3dsbh2}
\ee
where:
 $\eta_i=(1,\, -1,\, -1)$ and $\eta_i=(1,\, 1,\, -1)$  for $SO(2,4)$ and
$SO(4,2)$ supergravities, respectively.  
Note that the integration constants $\eta_i$ are
determined by  the Einstein equation
 $R_r^r+2\, R_\theta^\theta= -2\, V$.  (See also  section
3.2 of \cite{becvsa}.)
The conditions $H_i\ge 0$ ensure that the metric remains real, and that
the scalar fields are  in the physical regime $X_i\ge 0$. 
This requires $\mu\ge 0$, and it constrains the radial coordinate 
$r$  to lie only in a restricted range:
\be
\mu\ge 0\, , \qquad   0 \le  r^2\le  
\hbox{min}\left( \fft{\mu\sinh^2(\sqrt{k}\beta_2)}{k}\,,\,
\fft{\mu\sinh^2(\sqrt{k}\beta_3)}{k} \right) \, ,  \label{paramcons}
\ee
or
\be
\mu \ge 0\, , \qquad 0\le  r^2\le  
\fft{\mu\sinh^2(\sqrt{k}\beta_3)}{k} \, ,  \label{paramcons1}
\ee
for $SO(2,4)$ or $SO(4,2)$ gauged supergravity respectively.  Note
that the horizon(s) are determined by the zeros of the function $f$.
Their location depends on the values of the parameters $\mu$, $g$ and
$\beta_i$. (For a related discussion of horizons for AdS charged black
holes in $SO(6)$ gauged supergravity, see section 4.3 of
\cite{becvsa}.)  The solution has a curvature singularity both on the
lower and upper limit of the $r$ coordinate range (\ref{paramcons})
(for $SO(2,4)$ supergravity) and (\ref{paramcons1}) (for $SO(4,2)$
supergravity).

 The analytic continuation:
\be 
t\to \im\, t\, , \qquad  r\to \im\, r\, , \qquad  \theta \to \im\, \theta\,
\label{analcon} 
\ee
yields  a set of solutions  with  $\mu\le 0$, for which the $r$ coordinate
is restricted to a range: 
\be 
\mu\le 0\, ,\qquad  0 \le r^2 \le
\hbox{min}\left( \fft{|\mu|\sinh^2(\sqrt{k}\beta_2)}{k}\, ,\,
\fft{|\mu|\sinh^2(\sqrt{k}\beta_3)}{k}\right) \,, \label{cospar} 
\ee
or
\be 
\mu\le 0\, ,\qquad  0 \le r^2 \le
\fft{|\mu|\sinh^2(\sqrt{k}\beta_3)}{k}\, , \label{cospar1} 
\ee
for $SO(2,4)$ or $SO(4,2)$ gauged supergravity respectively.

   Owing to the analytic continuation performed in(\ref{analcon}), the
solution with $k=+1$ corresponds to a cosmological solution on ${\bf
H}^3$ with a ``big crunch'' singularity at the upper boundary of the
time coordinate $r$, and a cosmological horizon at a zero of function
$f=0$.  On the other hand the analytic continuation of the solution
with $k=-1$ corresponds to a black hole solution with $S^3$ sections, and a
naked singularity on both boundaries of the radial coordinate $r$.

   While the Ans\"atze for these solutions bear similarities to the
Ans\"atze for the $SO(6)$ AdS black hole solutions, the former
are highly singular.  This can be attributed to the fact
that the potential (\ref{d5pot}) does not have an extremum, and is
unbounded from above and below.

   The limit $\mu\to 0$ and $\beta_i\to \infty$, keeping $\mu\, e^{2
\beta_i}=2\, q_i$ finite, leads to supersymmetric
solutions.  For this class of solution one can perform an analytic
continuation $g\to \im \, g$ and formally solve the Killing spinor
equations with the new ansatz for the metric:\footnote{See
Ref. \cite{KlemmSabra,BehrndtCvetic} for details. These solutions are
analogues of four-dimensional charged de-Sitter solutions first
discussed by Kastor and Traschen \cite{KastorTraschen}.}
\bea 
ds_5^2 &=& -(H_1H_2H_3)^{-2/3}\, dt^2 + (H_1H_2H_3)^{1/3}\, (
dr^2 + r^2 d\Omega_{3,k=+1}^2)\,,\nn\\
X_i&=& H_i^{-1}(H_1H_2H_3)^{1/3}\, , \qquad A^i_\1 =\sqrt{k}\, (1-H_i^{-1})\,
 dt\,, \qquad\,, 
\label{d3dsPBS}
\eea
with new harmonic functions:
\be
H_i =2 \eta_i g\, t + \fft{q_i}{r^2}\, , 
\label{d3dsbhBPS1}
\ee
where
 $\eta_i=(1,\, -1,\, -1)$ and $\eta_i=(1,\, 1,\, -1)$   are solutions for 
 $SO(4,2)$ and
$SO(2,4)$ supergravities respectively.  (Note that the analytic continuation
$g\to \im\,
g$ changes the overall sign of the gauged supergravity potential, and thus
interchanges the $SO(4,2)$ and $SO(2,4)$ potentials.) 
This form of solutions allows for multi-centered  black hole solutions, i.e.:
\be
H_i = 2\eta_i g\, t + \sum_{j=1}^N\, \fft{q_{ij}}{({\vec r}_i-{\vec 
r})^2}\, . 
\label{d3dsbhBPS2}
\ee
Again, the positivity of the harmonic functions $H_i$ constrains the allowed
range of the $t$ and $r$ coordinates.

   The solution (\ref{d3dsPBS}) can be obtained (see
 \cite{KlemmSabra}) from the supersymmetric limit ($\mu\to 0$,
 $\beta_i\to \infty$, $\mu e^{2\beta_i}=2\, q_i$) of the solution
 (\ref{d3adsbh}), by first performing the analytic continuation $g\to
 \im \, g$ and the coordinate transformation
\be
r=r'\sqrt{2g \, t'}, \qquad \fft{d t'}{2g\,t'}=dt +F(r)dr, 
\qquad F(r) =\fft{-2g\, r
\prod_{i=1}^3 H_i(r)}{1-4g^2\,r^2\,  \prod_{i=1}^3 H_i(r)}\, ,\label{trans}
\ee
and then dropping the ``primes''.

\subsection{Four-charge solutions of  $D=4$, $N=2$ gauged $SO(4,4)$, $SO(2,6)$
and $SO(6,2)$ supergravities}

   In the $N=2$ truncation of four-dimensional $SO(4,4)$, $SO(2,6)$
and $SO(6,2)$ gauged supergravities, one can again find four-charge
solutions, whose ansatz is closely related to the four-charge
solutions of $SO(8)$ supergravity \cite{Duffliu,Cveticgubser}:
\bea 
ds_4^2&=&(H_1H_2H_3H_4)^{-1/2}(-f dt^2 )+ 
(H_1H_2H_3H_4)^{1/2}(f^{-1}\, dr^2
+r^2 d\Omega_{2,\, k}^2) \,
\\  \nn
X_i&=&
H_i^{-1}(H_1H_2H_3H_4)^{1/4}\, ,  \qquad A^i_\1 = \sqrt{k}\,
(1-H_i^{-1})\, \coth(\sqrt{k}\, \beta_i)\, dt\,, 
\eea
with
\be 
f=k-\fft{\mu}{r} +4g^2 r^2 H_1H_2H_3H_4\, ,
\ee
and  the harmonic functions
 \be
H_i = \eta_i +\fft{\mu\sinh^2(\sqrt{k}\beta_i)}{k r}\, , \qquad i=1,\cdots, 4
\,.
\label{har4d} 
\ee 
Here, one takes $\eta_i=(1,\, 1,\, -1,\, -1)$, $\eta_i=(1,\, -1,\,
-1,\, -1)$ and $\eta_i=(1,\, 1,\, 1,\, -1)$ for the $SO(4,4)$,
$SO(2,6)$ and $SO(6,2)$ gauged supergravities respectively.  The
constraints on the integration constants $\eta_i$ are again imposed by
the Einstein equations.

   The  conditions $X_i\ge 0$  imply positivity  of
harmonic functions $H_i$, and they constrain the parameter and the range
of the radial coordinate:
\bea 
&& \mu >0\, ,\qquad 0 \le r\le
\hbox{min}\!\!\left(\fft{\mu\sinh^2(\sqrt{k}\beta_3)}{k}\, ,
\fft{\mu\sinh^2(\sqrt{k}\beta_4)}{k} \right) \,,\\
&&\mu >0 \, ,\qquad 0 \le r\le
\hbox{min}\!\!\left(\fft{\mu\sinh^2(\sqrt{k}\beta_2)}{k}\, , 
\fft{\mu\sinh^2(\sqrt{k}\beta_3)}{k}\, ,
\fft{\mu\sinh^2(\sqrt{k}\beta_4)}{k} \right) \, ,\\
&&\mu >0 \, ,\qquad 0 \le r\le \fft{\mu\sinh^2(\sqrt{k}\beta_4)}{k}\,,
\eea
for the $SO(4,4)$, $SO(2,6)$ and $SO(6,2)$ gauged supergravities respectively.
These solutions have
naked singularities at both boundaries. 
Depending on the values of the $g$, $\mu$ and $\beta_i$ parameters, 
the solutions can have horizons at zeros of the function $f$.
There is also a mirror region with $\mu<0$ and $r<0$.

    After performing the analytic continuation $\mu\to \im\, \mu$, $r\to
\im\, r$, $t\to \im\, t$ and $\theta\to \im\, \theta$ of the solutions
with $k=1$, one obtains cosmological solutions on ${\bf H}^2$ that are
analogous to those in the five-dimensional case; they can have a a
``big crunch'' singularity at the upper bound of the time coordinate
$r$, and a cosmological horizon at a zero of the function $f$, whose
location is determined by the value of the parameters.

    In spite of the fact that the potential has a (tachyonic)
extremum, these solutions are singular, owing to the unbounded nature of
the potential.  

   The  limit 
 $\mu\to 0$ and $\beta_i\to \infty$, keeping $\mu e^{2 \beta_i}=2\, q_i$
finite,  leads to the   supersymmetric solutions, in a fashion analogous to
the $D=5$ case we discussed previously. 
In the present case,  an analytic continuation
$g\to \im \, g$  allows for a  class of solutions that take the form (see
\cite{BehrndtCvetic}):
\bea 
ds_5^2 &=& -(H_1H_2H_3H_4)^{-1/2}\, dt^2 + (H_1H_2H_3H_4)^{1/2}\, (
dr^2 + r^2 d\Omega_{3,k=+1}^2)\,,\nn\\
X_i&=& H_i^{-1}(H_1H_2H_3H_4)^{1/4}\, ,\qquad 
A^i_\1 =\sqrt{k}\, (1-H_i^{-1})\,
 dt\,,  
\label{d3dsPBS7}
\eea
with 
\be
H_i = 2\, \eta_i g\, t + \fft{q_i}{r}\, .
\label{d4dsbhBPS}
\ee
 Here
 $\eta_i=(1,\, 1,\, -1,\, -1)$, $\eta_i=(1,\, -1,\, -1,\, -1)$ and   
$\eta_i=(1,\,
 1,\, 1, \, -1)$  for
 $SO(4,4)$,
$SO(6,2)$ and $SO(2,6)$ gauged supergravities, respectively.  
(Again the analytic continuation $g\to \im \, 
g$ changes the overall sign of the gauged supergravity potential and thus
interchanges the $SO(6,2)$ potential with the $SO(2,6)$ one.) 
These solutions  also allow for 
 multi-centered  black holes, i.e.
\be
H_i = 2\, \eta_i g\, t + \sum_{j=1}^N\, \fft{q_{ij}}{|{\vec r}_i-{\vec 
r}|}\, . 
\label{d4dsbhBPS1}
\ee
Again, the positivity of the harmonic functions $H_i$ constrains the allowed
range of the $t$ and $r$ coordinates.

   These solutions are different from the four-charge de Sitter
solutions of stable de Sitter vacua, discussed in
\cite{BehrndtCvetic}. The latter can be obtained by the analytic
continuation $g\to \im g$ of the BPS four-charge black hole solutions
in $SO(8)$ gauged supergravity.  They correspond to the four-charge
solutions (\ref{d3dsPBS7})-(\ref{d4dsbhBPS1}) with $\eta_i=(1,\, 1\,\,
1,\, 1)$ and asymptote at late times to the stable de Sitter vacuum.
When one identifies the four charges $q_i=q$ ($i=1,\cdots 4$) the
scalars become constant ($X_i=1$) and the solution becomes the de
Sitter Reissner Nordstr\"om black hole of Kastor and Traschen
\cite{KastorTraschen} of the (stable) de Sitter vacuum.

    In the case of $SO(4,4)$ gauge supergravity the potential also has
a tachyonic de Sitter extremum. However, the solutions
(\ref{d3dsPBS7})-(\ref{d4dsbhBPS1}) with $\eta_i=(1,\, 1, \, -1,\, -
1)$ do not admit a limit of the charge parameters $q_i$ for which the
scalars could become constant ($X_i=1$), and thus these solution cannot
describe charged de Sitter black hole solutions in this tachyonic
extremum.

\subsection{Two-charge solutions of  $D=7$, $N=2$ 
gauged $SO(1,4)$ and $SO(3,2)$ supergravities}

   Our last examples arise in the $N=2$ truncation of
seven-dimensional $SO(1,4)$ and $SO(3,2)$ supergravities.  These
solutions are closely related to the AdS black hole solutions of the
$N=2$ truncation of seven-dimensional gauged $SO(5)$ supergravity. The
solutions here are of the form
\bea 
ds_7^2 &=&
-(H_1H_2)^{-4/5}\, f\, dt^2 + (H_1H_2)^{1/5}\, (f^{-1}\, dr^2 + r^2
d\Omega_{5,k}^2)\ ,\\ \nn  
X_i&=& H_i^{-1}(H_1H_2)^{1/2}\, , (i=1,\, 2)\, , 
 \qquad
X_0=(X_1X_2)^{-2}\, , \nn\\
A^i_\1 &=& \sqrt{k}\, (1-H_i^{-1})\, \coth(\sqrt{k}\,
\beta_i)\, dt\,,\qquad i=1,2\,, \label{d2dsbh} 
\eea 
with 
\be 
f =k-\fft{\mu}{r^4} + 4g^2\, r^2\, (H_1H_2) \, ,  \label{sc2ds} 
\ee
and  harmonic functions given by
\be 
H_i = \eta_i + \fft{\mu\, \sinh^2(\sqrt{k}\, \beta_i)}{k\, r^4}\,.
\label{d2dsbh2} 
\ee 
Here, $\eta_i=(-1, \, -1)$ and $\eta_i=(1,\, -1)$ for the $SO(1,4)$
and $SO(3,2)$ gauged supergravities respectively. The positivity of
the $X_i$ is ensured by requiring the harmonic functions $H_i$ to be
positive, which is achieved by the restrictions
\be 
\mu\ge 0\, , \qquad  0\le r^4 \le
\hbox{min}\left(\fft{\mu\sinh^2(\sqrt{k}\beta_1)}{k}\, ,\,
\fft{\mu\sinh^2(\sqrt{k}\beta_2)}{k} \right) \,,\label{paramcons2}
\ee
or
\be 
\mu\ge 0\, , \qquad  0\le r^4 \le
 \fft{\mu\sinh^2(\sqrt{k}\beta_2)}{k} \,,\label{paramcons3}
\ee
respectively. The analytic continuation 
\be 
t\to \im\, t\, , \qquad  r\to \im\, 
r\, , \qquad \theta \to \im\, \theta\, \label{analcon2} 
\ee
yields a cosmological solution with  $\mu\le 0$, and the time coordinate $r$
is constrained to the same region as above. 

    Taking the supersymmetric limit and analytic continuation $g\to
\im \, g$, one also obtains analogous ``de Sitter'' solutions:
\bea 
ds_7^2 &=&
-(H_1H_2)^{-4/5}\, dt^2 + (H_1H_2)^{1/5}\, ( dr^2 + r^2
d\Omega_{5,k=+1}^2)\ ,\\ \nn  X_i&=& H_i^{-1}(H_1H_2)^{1/2}\, , \qquad
A^i_\1 = \sqrt{k}\, (1-H_i^{-1}) \,
\beta_i)\, dt\,,\qquad i=1,2\,, \label{d2dsbhBPS} 
\eea 
with the  harmonic functions
\be 
H_i = 2\, \eta_i g\, t + \fft{q_i}{ r^4}\,.
\label{d2dsbh2BPS} 
\ee 

   We conclude this section with a comment about another solution
associated with the $N=2$ truncation of seven-dimensional supergravity, 
with the $SO(2,2)$ gauging.  The domain wall solution \cite{Sezginetal} of
the $N=2$ supergravity for the $SO(4)$ gauging is given by
\be
ds_7^2=e^{2A}\, dx^\mu\, dx^\nu \eta_{\mu\,\nu}+e^{8A}\, dy^2\, ,\qquad
e^{-\textstyle{{3}\over{\sqrt{10}}}\phi} =H\,,
\ee
with
\be 
e^{-4\, A}= {{2g}\over {5(H^{-1/3})'}}\, , \qquad H=
e^{-\textstyle{{3}\over{\sqrt{10}}}\phi_0} +q|y|\,, 
\ee
where a prime denotes a derivative with respect to $y$.  The analytic
continuations $g\to \im\, g$ and $\{t,\, y,\, q\}\to \im\, \{t, \,
y,\, q\}$ turn this into a cosmological solution of the de Sitter
gauged supergravity.

\section{Conclusions}

   An elegant feature of gauged $SO(p,q)$ ($q=2r$) supergravity
theories is that they can be obtained from gauged $SO(p+q)$
supergravity theories by means of straightforward analytic
continuations. The consistent Pauli reductions of M-theory and string theory on
spheres $S^{p+q-1}$, yielding gauged $SO(p+q)$ supergravities, have
been studied extensively in the literature. This has allowed us to explore
in a straightforward way the corresponding consistent Pauli reductions
on the hyperboloidal spaces ${\cal H}^{p,q}$, thus yielding 
supergravity theories with non-compact gauge groups.

   We provided a general analysis of the extrema of the unimodular
part of the scalar potential for supergravities with $SO(p,q)$
gaugings. It turns out that only for seven-dimensional supergravity with
an $SO(2,2)$ gauging, and four-dimensional supergravity with an $SO(4,4)$
gauging, does one obtain stable extrema of the unimodular part of the scalar
potential. In the seven-dimensional case the potential still depends
on a ``volume'' scalar, thus yielding (cosmological) solutions with
a running scalar and a positive potential contribution.  On the
other hand, the four-dimensional case has an extremum of the scalar
potential corresponding to a de Sitter vacuum, which, however, is a
{\it saddle point}.  This result is consistent with the analysis of
the $N=2$ gauged supergravity potentials with general non-compact
isometries \cite{Kallosh}.

   Interestingly, we also found that $D=4$ gauged $SO(4,4)$, $SO(2,6)$
and $SO(6,2)$ supergravity, $D=5$ gauged $SO(2,4)$ and $SO(4,6)$
supergravity and $D=7$ gauged $SO(1,4)$ and $SO(3,2)$ supergravity
admit Abelian charged black hole (and cosmological) solutions in their
$N=2$ truncations, whose structures are closely related to the
corresponding gauged $SO(p+q)$ supergravity black holes.  However, the
solutions obtained here are highly singular, in consequence of the
unbounded nature of the scalar potentials.

\section*{Acknowledgments}

We should like to thank Klaus Behrndt, Hong L\"u and Toine van Proeyen
for discussions. We should also like to thank the Benasque workshop,
Cambridge workshop (M.C. and C.N.P.), CERN (M.C.), UPenn (G.W.G,
C.N.P.) and Mitchell Institute (M.C. and G.W.G) for hospitality and 
support during the course of this work.

\end{document}